\documentclass[twocolumn,epj]{svjour}  
\smartqed  % flush right qed marks, e.g. at end of proof

\usepackage{longtable}
\usepackage{grffile} %deal with fig name with ".,_ " .etc...
\usepackage{graphics}
\usepackage{graphicx} %set figure size with width=...
\usepackage{amsmath}    % need for subequations
\usepackage{hyperref}   % use for hypertext links, including those to external documents and URLs
\usepackage{subfigure}  % use for figure side by side
\usepackage{epsfig}
\usepackage{epstopdf}
\usepackage{color}
\usepackage{url}

\usepackage{lineno}
\begin{document}

\title{Investigating high energy proton proton collisions with a multi-phase transport model approach based on PYTHIA8 initial conditions}

\author{Liang Zheng\inst{1,4}  \thanks{{e-mail:} zhengliang@mail.cug.edu.cn}
	\and Guang-Hui Zhang\inst{1} \and Yun-Fan Liu \inst{1}
	\and Zi-Wei Lin\inst{2} \thanks{{e-mail:} linz@ecu.edu}
	\and Qi-Ye Shou\inst{3} \thanks{{e-mail:} shouqiye@fudan.edu.cn}
	\and Zhong-Bao Yin\inst{4} \thanks{{e-mail:} zbyin@mail.ccnu.edu.cn}
}

\institute{School of Mathematics and Physics, China University of Geosciences (Wuhan), 
	Lumo Road 388, Wuhan 430074, China 
	\and
Department of Physics, East Carolina University,
Greenville, North Carolina 27858, USA
	\and
	Key Laboratory of Nuclear Physics and Ion-beam Application (MOE),
	Institute of Modern Physics, Fudan University, Shanghai 200433, China
	\and
Key Laboratory of Quark and Lepton Physics (MOE) and Institute
of Particle Physics, Central China Normal University, Wuhan 430079, China
}

\date{\\}%Received: \today / Revised version: \today}

\abstract{ The striking resemblance of high multiplicity proton-proton (pp) collisions at the LHC to heavy ion collisions challenges our conventional wisdom on the formation of the Quark-Gluon Plasma (QGP). A consistent explanation of the collectivity phenomena in pp will help us to understand the mechanism that leads to the QGP-like signals in small systems. In this study, we introduce a transport model approach connecting the initial conditions provided by PYTHIA8 with subsequent AMPT rescatterings to study the collective behavior in high energy pp collisions. The multiplicity dependence of light hadron productions from this model is in reasonable agreement with the pp $\sqrt{s}=13$ TeV experimental data. It is found in the comparisons that both the partonic and hadronic final state interactions are important for the generation of the radial flow feature of the pp transverse momentum spectra. The study also shows that the long range two particle azimuthal correlation in high multiplicity pp events is sensitive to the proton sub-nucleon spatial fluctuations.
%\keywords{Transport model \and Collectivity \and High multiplicity pp collision}
}

\authorrunning{Liang Zheng et. al.}
\titlerunning{Investigating high energy pp collisions with a multi-phase transport model approach}

\maketitle

\section{Introduction}
\label{intro}
Collective phenomena have been regarded as important signatures indicating the
creation of a new state for nuclear matter, Quark-Gluon Plasma (QGP), in
relativistic heavy ion (AA) collisions~\cite{Broniowski:2008vp,Gale:2012rq,Noronha-Hostler:2015uye}. In recent years, a flood of striking
collectivity like features have also been observed in high multiplicity
proton-proton (pp) and proton-nucleus (pA) collisions at the Relativistic Heavy-Ion Collider and the Large Hadron Collider (LHC)~\cite{Dusling:2015gta,Nagle:2018nvi}. These
observations were mostly unanticipated for small systems like pp which were
expected to be QGP free references for the study of the de-confined
quark gluon matter produced in AA collisions. An early example like this comes
from the observed ridge structure of the azimuthal correlations of two charged
hadrons with a large pseudo-rapidity gap~\cite{Khachatryan:2010gv,Aad:2015gqa,Khachatryan:2016txc}. Later
discoveries on the enhancement of multi-strange particles~\cite{ALICE:2017jyt,Acharya:2019kyh,Adam:2015vsf}, non-vanishing elliptic
flow coefficients~\cite{Khachatryan:2016txc,Acharya:2019vdf,Abelev:2014mda,Aaboud:2017blb}, mass dependence of the hadron $p_T$ spectra\cite{Adam:2015qaa,Acharya:2020zji,Acharya:2018orn,Abelev:2014qqa} and
characteristic variations of the $p_T$ dependent baryon to meson ratios~\cite{Acharya:2018orn,Abelev:2013haa,Acharya:2020zji,Khachatryan:2011tm} further
consolidate the resemblance between the small system and AA events.

The observation of the surprising collectivity in pp collisions challenges our conventional wisdom on the formation of the QGP medium and invites a lot of theoretical studies to understand its origin~\cite{Nagle:2018nvi,Adolfsson:2020dhm,He:2015hfa,Lin:2015ucn}. 
It is conceived that a QGP droplet can be partly formed in high multiplicity pp events and experience the same hydrodynamic expansion as in AA collisions~\cite{Qin:2013bha,Bzdak:2013zma,Habich:2015rtj,Weller:2017tsr,Zhao:2017rgg,Zhao:2020pty}. Similar ideas are also implemented with a core-corona picture in the EPOS model~\cite{Pierog:2013ria,Werner:2013tya,Kanakubo:2019ogh}. Alternative approaches focusing on the initial state gluon correlations based on the color glass condensate are also found to be successful in describing the qualitative features of the seemingly collective behavior~\cite{Dusling:2012iga,Dusling:2012cg,Schenke:2016lrs,Dusling:2017dqg}. 

Another group of works relies on extensions to the traditional string fragmentation model in the PYTHIA event generator~\cite{Sjostrand:2006za,Sjostrand:2014zea} by considering interactions between multiple string objects when they overlap in the transverse space. The color reconnection model revises the string structure created in the events with multiple parton interactions (MPI) and is useful in dealing with the flavor dependent observables~\cite{Ortiz:2013yxa,Bierlich:2015rha}. The color rope model, initially applied to pp collisions in DIPSY~\cite{Bierlich:2014xba}, now a part of the standard PYTHIA program, is also capable to reasonably describe the multiplicity dependent particle compositions~\cite{Bierlich:2017sxk}. However, dynamical correlations, like the near side ridge and elliptic flow observables, are generally difficult to reproduce within the traditional string fragmentation framework. The recent development of the string shoving mechanism~\cite{Bierlich:2016vgw,Bierlich:2017vhg,Bierlich:2020naj} is expected to improve the description of these measurements by considering the repelling of string pieces overlapped in space-time. 

On the other hand, the string melting version of a multi-phase transport model (AMPT)~\cite{Lin:2004en}, which includes both partonic and hadronic final state rescattering effects, has been extensively applied to interpret the collective flow both in AA and in small systems~\cite{Bzdak:2014dia,Bozek:2015swa,Nie:2018xog,Nagle:2017sjv,Shao:2020sqr,Ma:2016fve}. With the event by event geometry fluctuations, this method provides a generic way to understand the importance of the parton degrees of freedom in collectivity observables and shows the non-equilibrium effect such as the parton escape mechanism can be quite significant for the development of collective flow especially in small systems~\cite{He:2015hfa,Lin:2015ucn,Li:2016ubw}. Whilst being successful in depicting the azimuthal anisotropic flow, the AMPT model can not easily reproduce the multiplicity dependence of transverse momentum spectra until the latest improvement~\cite{Zhang:2021vvp}. In this work, we build a transport model framework based on the initial conditions provided by PYTHIA8 linked to the final state interactions and quark coalescence model~\cite{He:2017tla} within AMPT, in an effort to simultaneously describe the $p_T$ spectra and multi-particle correlations in pp collisions at $\sqrt{s}=13$ TeV. One can exploit the interplay of parton and hadron final state interactions by considering the microscopic dynamical process of the evolving system. This study may pave the way for further investigations on the azimuthal anisotropy measurements in small systems and help us to fathom their origins from the perspective of a transport model.  

The rest of this paper is organized as follows: we elaborate the 
transport model approach for pp collisions and its extensions with the proton transverse spatial geometries in Sec.~\ref{sec:model}. Comparisons of the model calculations to the multiplicity dependent pp data from the LHC are made in Sec.~\ref{sec:results}. Finally, we summarize our major conclusions and discuss the potential applications of this framework in Sec.~\ref{sec:summary}.

\section{Model setup}
\label{sec:model}

Measurements in small collision systems have shown anisotropic flows similar to those observed in heavy ion collisions in which hydrodynamic descriptions are found to work well~\cite{Schenke:2021mxx}. This implies that a deconfined quark gluon matter may have been created in these small systems. In the PYTHIA framework, the partonic system including jets and perturbative radiations is kept in the strings even in the high energy density region, which might underestimate the collective partonic effects. To account for the effects of the parton degrees of freedom, we adopt the string melting mechanism from AMPT, converting excited strings generated by PYTHIA to their constituent partons, and couple the resultant parton system to further final state rescatterings based on the kinetic transport model approach. 

A high energy pp event in this approach is described by the same four ingredients as the AMPT model: initial conditions, partonic rescatterings, hadronization and hadronic rescatterings. Unlike the full AMPT model which uses HIJING~\cite{Wang:1991hta} to obtain initial conditions, PYTHIA8 is utilized to generate the initial conditions for the subsequent evolution stages. The PYTHIA/Angantyr model~\cite{Bierlich:2018xfw} implemented in PYTHIA8 can provide an impact parameter dependent string system within the MPI framework suitable for further treatment considering its space-time structure. We consider the primary hadrons from the excited strings as the intermediate step to model the string melting process. The parton system is obtained by converting the primary hadrons to their valence quarks~\cite{Lin:2001zk}. A formation time 
\begin{equation}
t_{p}=E_{H}/m^{2}_{T,H}
\label{eqn:tprod}
\end{equation}
 is given to each primary hadron including its valence quarks, with $E_{H}$ and $m_{T,H}$ being the energy and transverse mass of its mother hadron. Partons do not have any interactions until after their formation time. The primary hadrons can be regarded as the parton sources, which give the parton initial space-time configuration after the propagation of the primary hadrons to their formation times. The microscopic dynamic processes of the parton evolution are implemented with the Zhang's Parton Cascade (ZPC) model~\cite{Zhang:1997ej} by solving the Boltzmann equation of two-body elastic scattering process through the parton cascade method. The parton rescattering cross section in ZPC determines the evolution feature of the deconfined parton medium. For the partons without further scatterings in ZPC, they will become hadrons through a spatial quark coalescence model. The hadron species are determined by the coalesced (anti)quark flavor combination with an overall parameter $r_{BM}=0.53$ dictating whether a meson or a baryon is formed~\cite{He:2017tla}. Further hadronic scatterings are treated with an extended relativistic transport model (ART)~\cite{Lin:2004en,Li:1995pra} with both elastic and inelastic scatterings. 

The key model parameters used in this approach can be divided into two categories, the parameters involved in the initial conditions and those relevant for the final state rescatterings. As the initial conditions are generated by PYTHIA8 in this framework, we take the Monash tune of PYTHIA8 parameters in the simulation. The key final state interaction parameter is the parton rescattering cross section in ZPC. Its value is determined by systematic comparisons with the elliptic flow data. The full AMPT model with the new quark coalescence uses 1.5 mb for large systems~\cite{He:2017tla}, while in this work we provide results with 0 mb, 0.2 mb and 3 mb in the comparison to demonstrate the parton rescattering effects based on our current approach. If we turn off all the final state interactions, the result is expected to be very similar to PYTHIA8 Monash calculations. 

\begin{figure*}[hbt!]
	\centering
	\includegraphics[width=4.5cm]{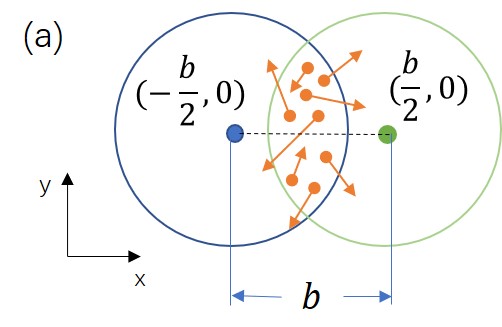}
	\includegraphics[width=5.2cm]{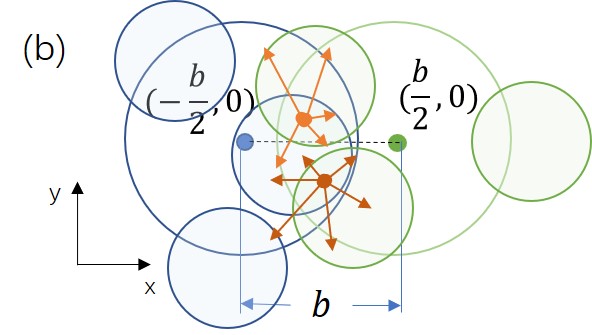}
	\includegraphics[width=4.8cm]{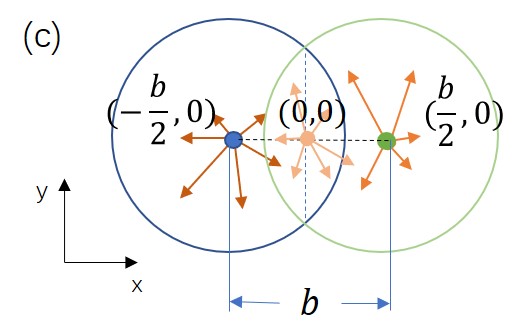}
	\caption{An illustration of the spatial distributions of the parton sources in the transverse plane produced with our transport model approach using the overlap function weighting method (a) and the constituent quark method (b), together with the distributions from the full AMPT model (c). Filled green and blue circles represent the projectile and target proton positions at impact parameter $b$, respectively. The two big green and blue circles show the transverse size of the incoming protons. Arrows represent the velocity directions with their attached bulbs indicating positions of the produced parton sources. Shaded circles in (b) are the constituent quarks from the projectile and the target. }
	\label{fig:trans_space}  
\end{figure*}

\begin{figure*}
	\centering
	\includegraphics[width=0.45\textwidth]{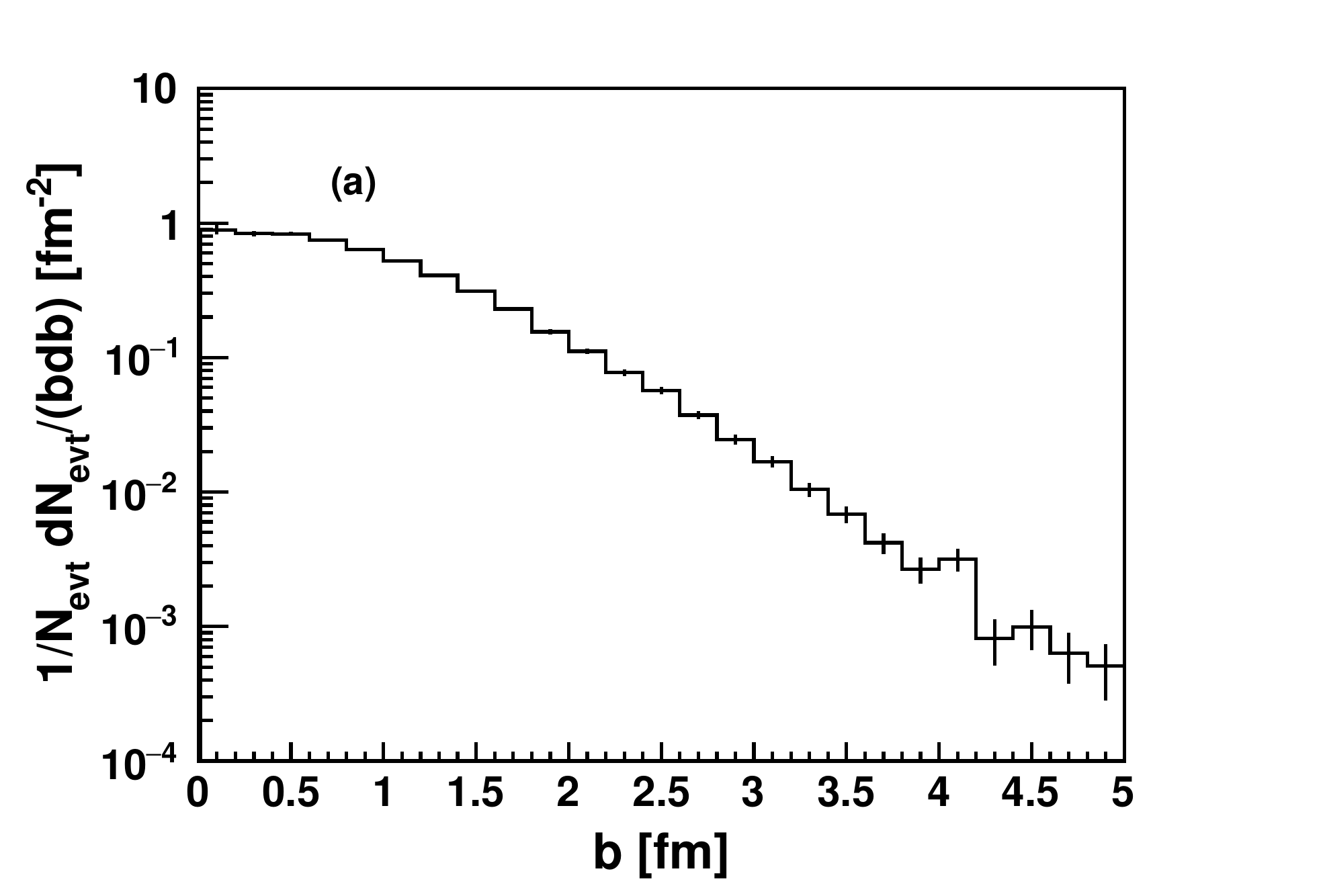}
	\includegraphics[width=0.45\textwidth]{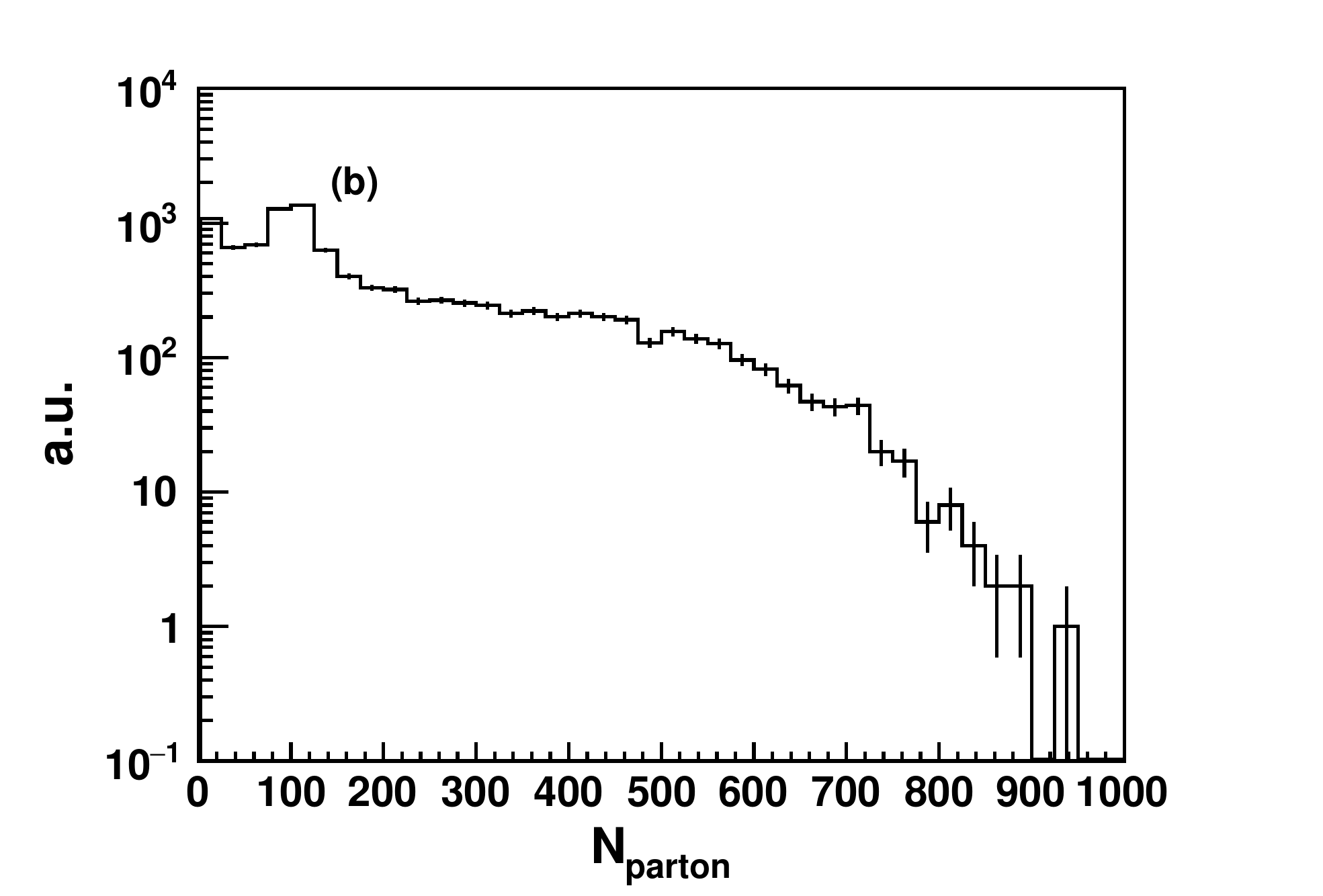}
	\caption{Normalized impact parameter $b$ distribution from PYTHIA/Angantyr (a) and the parton number distributions after string melting (b) for pp collisions at $\sqrt{s}=13$ TeV.}
	\label{fig:b_Npart}  
\end{figure*}

Subnucleonic geometry has been speculated to be important in the space-time development of the evolving nuclear matter produced in small collision systems. 
In our transport model approach, we take the sub-nucleon geometry into consideration when sampling the initial transverse positions of the parton sources before converted to constituent quarks. The sub-nucleon geometry may reveal itself together with strong spatial fluctuations in small system collective flow~\cite{Schenke:2021mxx,Schenke:2014zha,Mantysaari:2016ykx,Welsh:2016siu,Mantysaari:2017cni}. A precise constraint on the underlying parton spatial distribution of the nucleon might require the data from future deep inelastic scattering experiments~\cite{Accardi:2012qut}. We perform our study with two initial geometry models representing our current understanding of the proton inner structure.

\begin{figure*}[hbt!]
	\centering
		\includegraphics[width=0.45\textwidth]{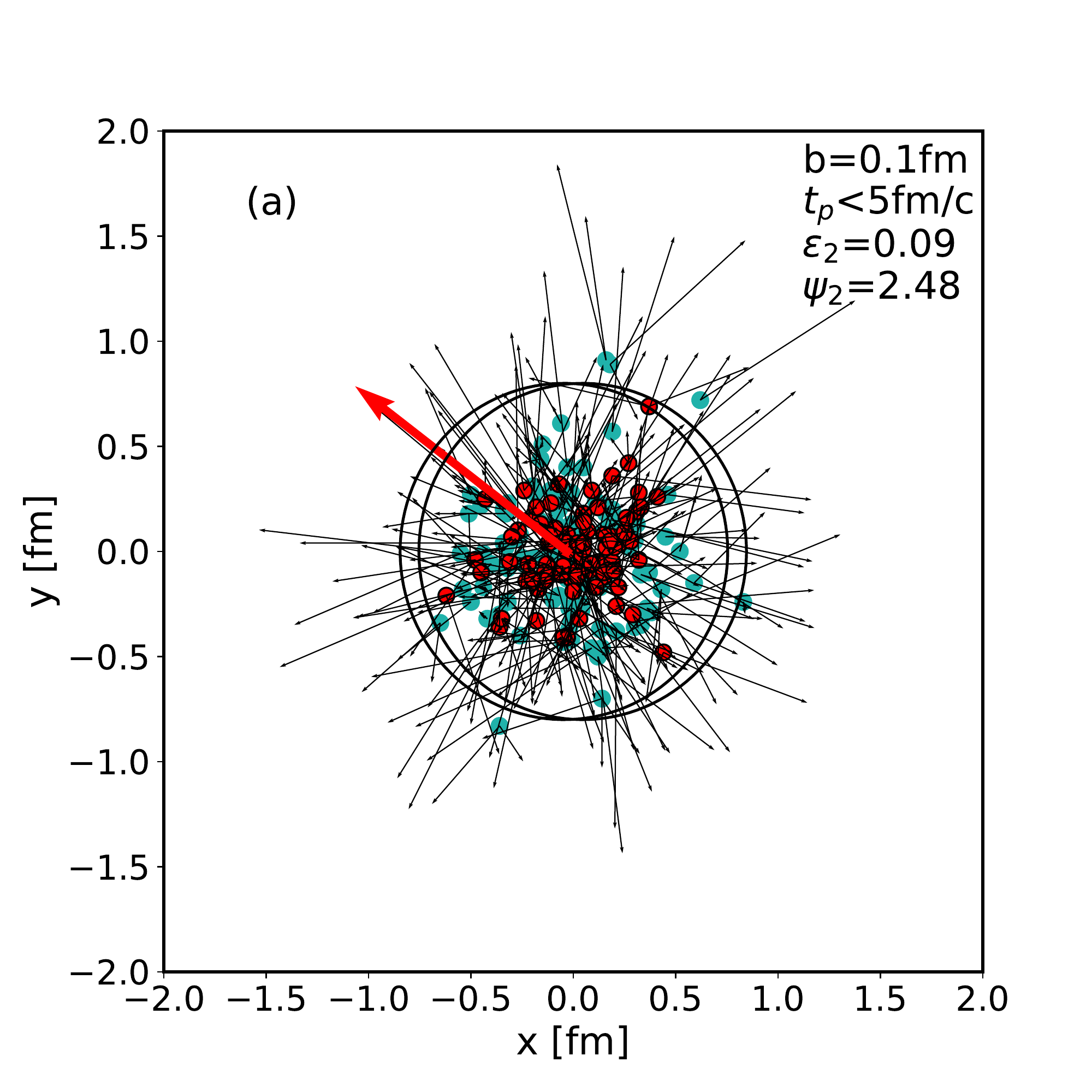}
		\includegraphics[width=0.45\textwidth]{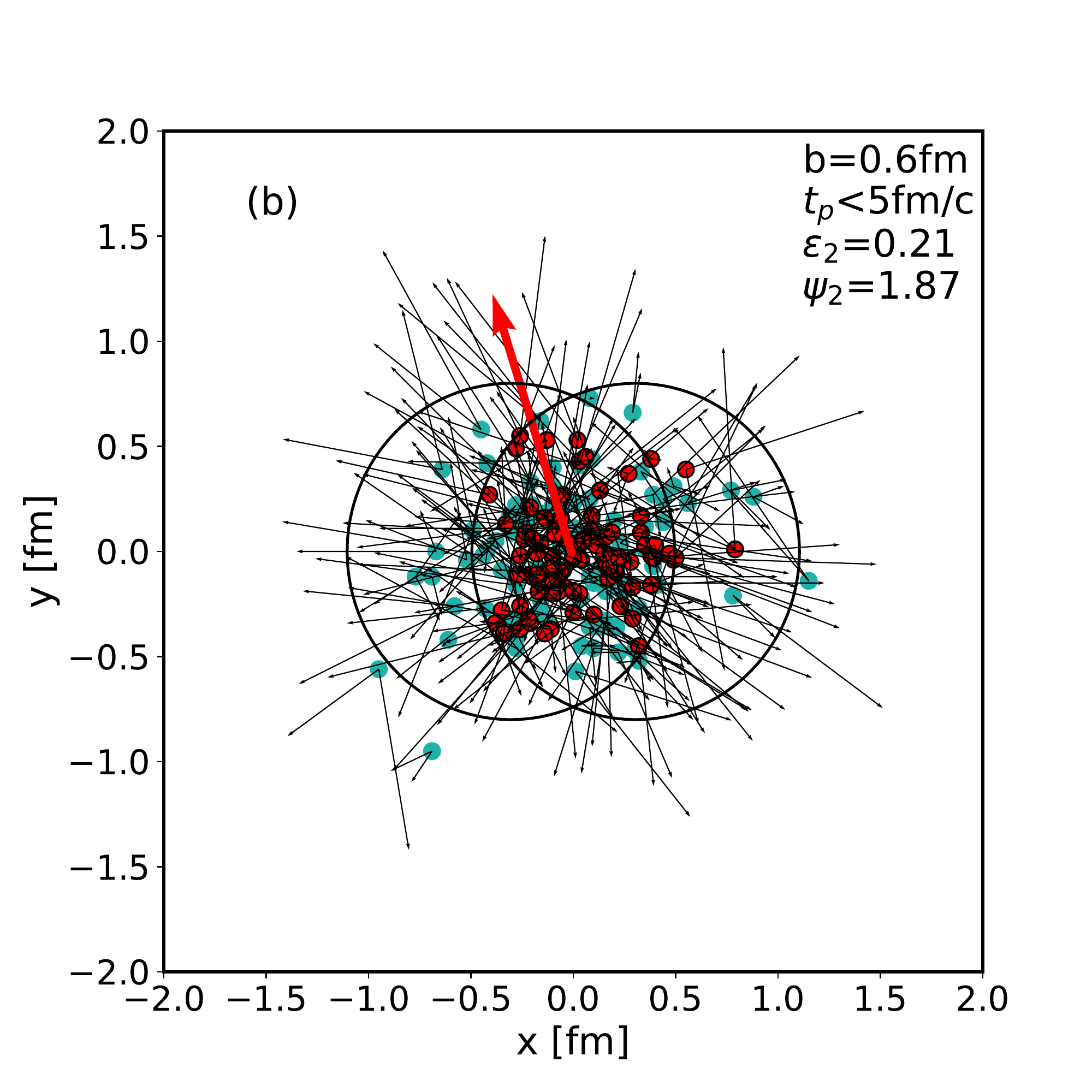}
		\includegraphics[width=0.45\textwidth]{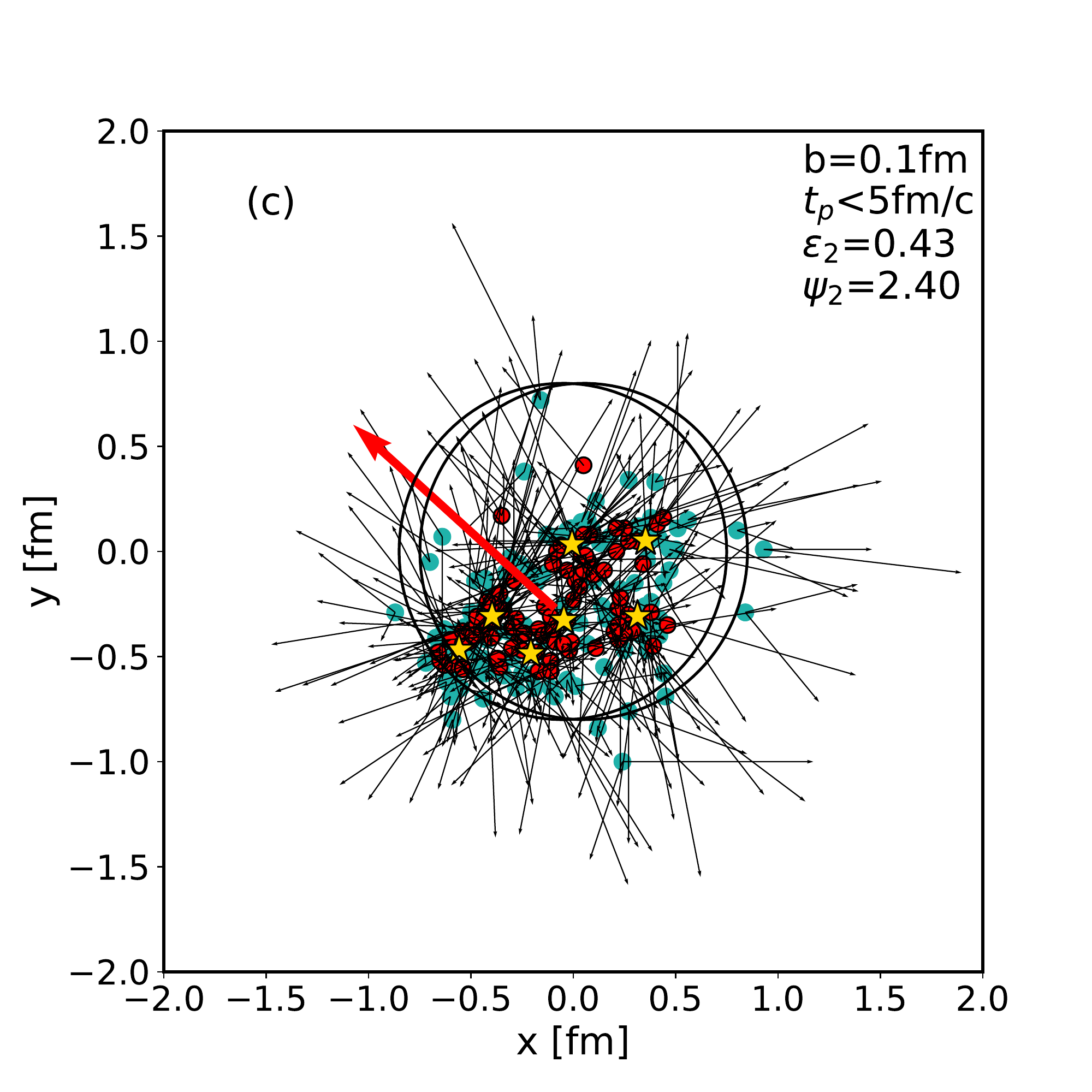}
		\includegraphics[width=0.45\textwidth]{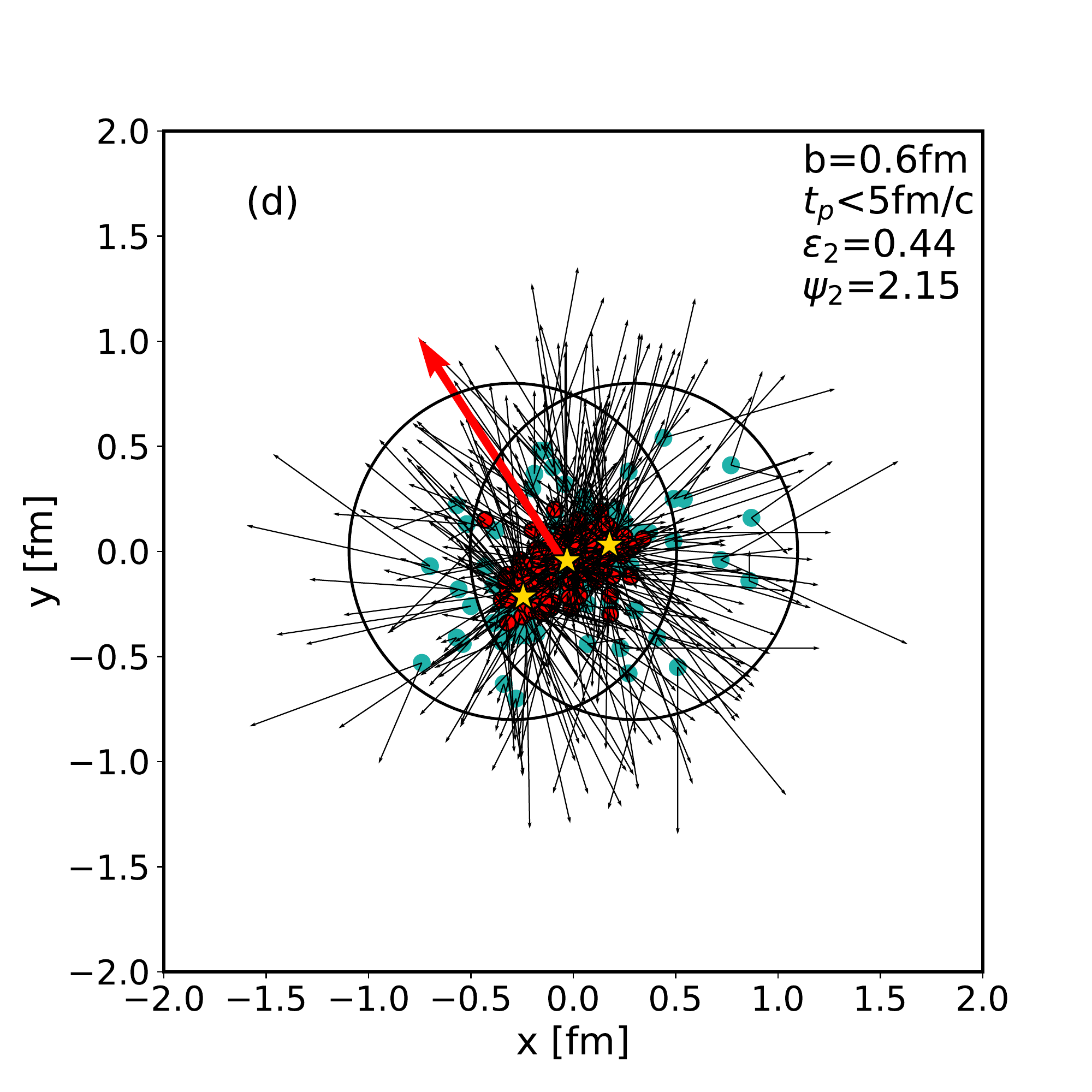}
	\caption{Transverse view of the created parton system in a high multiplicity event ($N_{parton}>600$) with the overlap function weighting method (upper panel) and the constituent quark method (lower panel) in a central collision at $b=0.1$ fm (left column) and in a peripheral collision at $b=0.6$ fm (right column). Coordinates of the partons are represented by the solid dots. Positions of the binary constituent quark collisions are shown by the yellow stars in (c) and (d). Initial parton velocity vectors are shown as black thin arrows. The red thick arrow shows the direction of the second order event plane.  }
	\label{fig:geo}       
\end{figure*}

Different matter distributions for proton have been extensively compared in~\cite{dEnterria:2010xip} and considered to be strongly connected to the eccentricity of the evolving system produced in pp events. 
It is common to parameterize the proton spatial density $\rho(r)$ with the following different distributions: hard sphere, exponential, Fermi, Gaussian and double-Gaussian functions. We provide results for the exponential matter distribution based on the proton charge form factor
\begin{equation}
\rho(r)=\frac{1}{8\pi R^3}e^{-r/R},
\label{eqn:exp_density}
\end{equation}
where $R=0.2$ fm. The transverse coordinates of the produced string objects are determined according to the overlap function of a pp collision at an impact parameter $b$ with $T(x, y, b)=\int\rho_{p,1}(x-b/2, y, z) \rho_{p,2}(x+b/2, y, z) dz$ assuming the two colliding protons are moving along the $z$ axis. A schematic illustration of the spatial distribution of the produced objects based on the overlap function $T(x,y,b)$ weighting method is provided in Fig.~\ref{fig:trans_space} (a).

The previous method gives a smooth initial transverse spatial condition but may underestimate the fluctuations in the proton matter distribution. We also introduce a Glauber Monte Carlo type method considering the event by event sub-nucleon level fluctuations based on the constituent quark picture~\cite{Zheng:2016nxx,Loizides:2016djv,Bozek:2016kpf,Bozek:2019wyr}. By assuming that a proton can carry three constituent quarks, the interaction of two protons can be extended to the participant quark geometries within the Glauber model framework. The constituent coordinates in a proton can be sampled according to the proton form factor as shown in Eq.~\ref{eqn:exp_density}. Two constituents from the different incoming protons thus may collide if their transverse separation is less than $D=\sqrt{\sigma_{cc}/\pi}$, with $\sigma_{cc}$ being the constituent scattering cross section. The transverse coordinates of the parton sources are randomly assigned to the binary collision center of each interacted constituent pair, as shown in Fig.~\ref{fig:trans_space} (b). The results shown in this work are obtained with $\sigma_{cc}=25.2$ mb, which reproduces the inelastic pp cross section at $\sqrt{s}=13$ TeV~\cite{Loizides:2016djv}. It must be emphasized that the constituent quark picture used in our current approach is different from an initial state quark participant Glauber Monte Carlo model~\cite{Bozek:2019wyr}. We first take the interacting nucleon collision geometries provided by PYTHIA/Angantyr for pp and then assign the produced particle transverse positions using the colliding quark spatial conditions as an afterburner to account for the event by event sub-nucleon fluctuations. 

As a comparison, the initial transverse spatial distributions of the parton sources produced in AMPT is illustrated in Fig.~\ref{fig:trans_space}(c). The parton system produced in AMPT based on the HIJING initial conditions can be divided into two types. The first type coming from the wounded projectile or target proton is attached to the position of the incoming projectile ($b/2,0$) or target ($-b/2,0$) beam in transverse plane, respectively. The second type from the possible minijet production process is put at the midpoint (0,0) of two beam locations. Unlike the approach in our current work modeling the parton spatial distributions based on the proton geometry in a dynamical way, the full AMPT model is placing the initial parton objects at three fixed locations largely dependent on the size of the impact parameter.

In kinetic transport model methods, the final state particle momentum anisotropy is converted from the initial spatial structures via the final state rescattering process. The initial spatial structure of an event is tightly related to the impact parameter distribution and the parton number created in the evolving system. We present the impact parameter distribution taken from PYTHIA/Angantyr for pp 13 TeV collisions in Fig.~\ref{fig:b_Npart}(a). It can be observed in this figure that the maximum impact parameter may rise up to 5 fm. The parton number distribution created after string melting process over the full phase space is shown in Fig.~\ref{fig:b_Npart}(b). The number of created partons in pp collisions at the LHC energy can be very large.

\begin{figure*}[hbt!]
	\centering
	\includegraphics[width=0.49\textwidth]{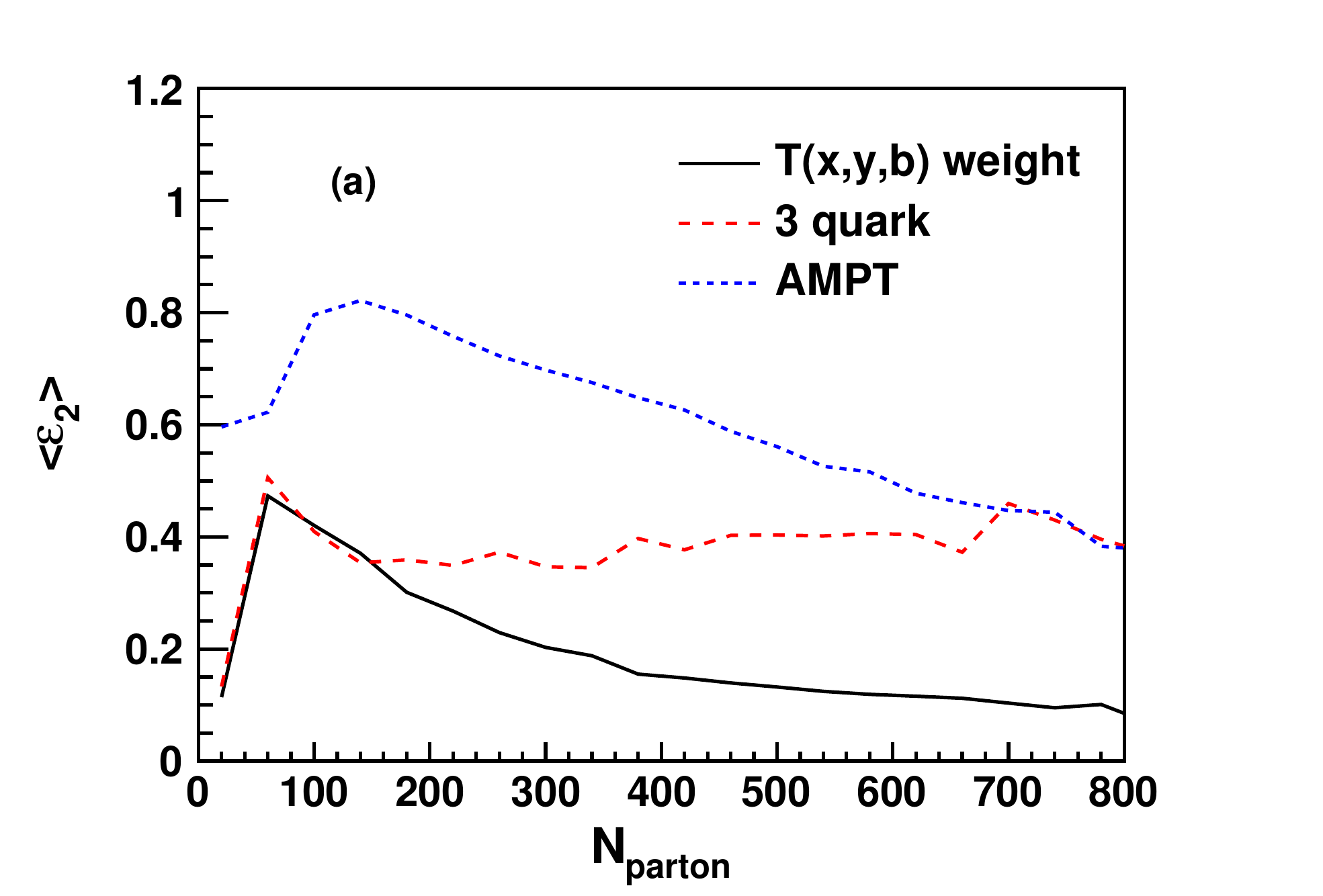}
	\includegraphics[width=0.49\textwidth]{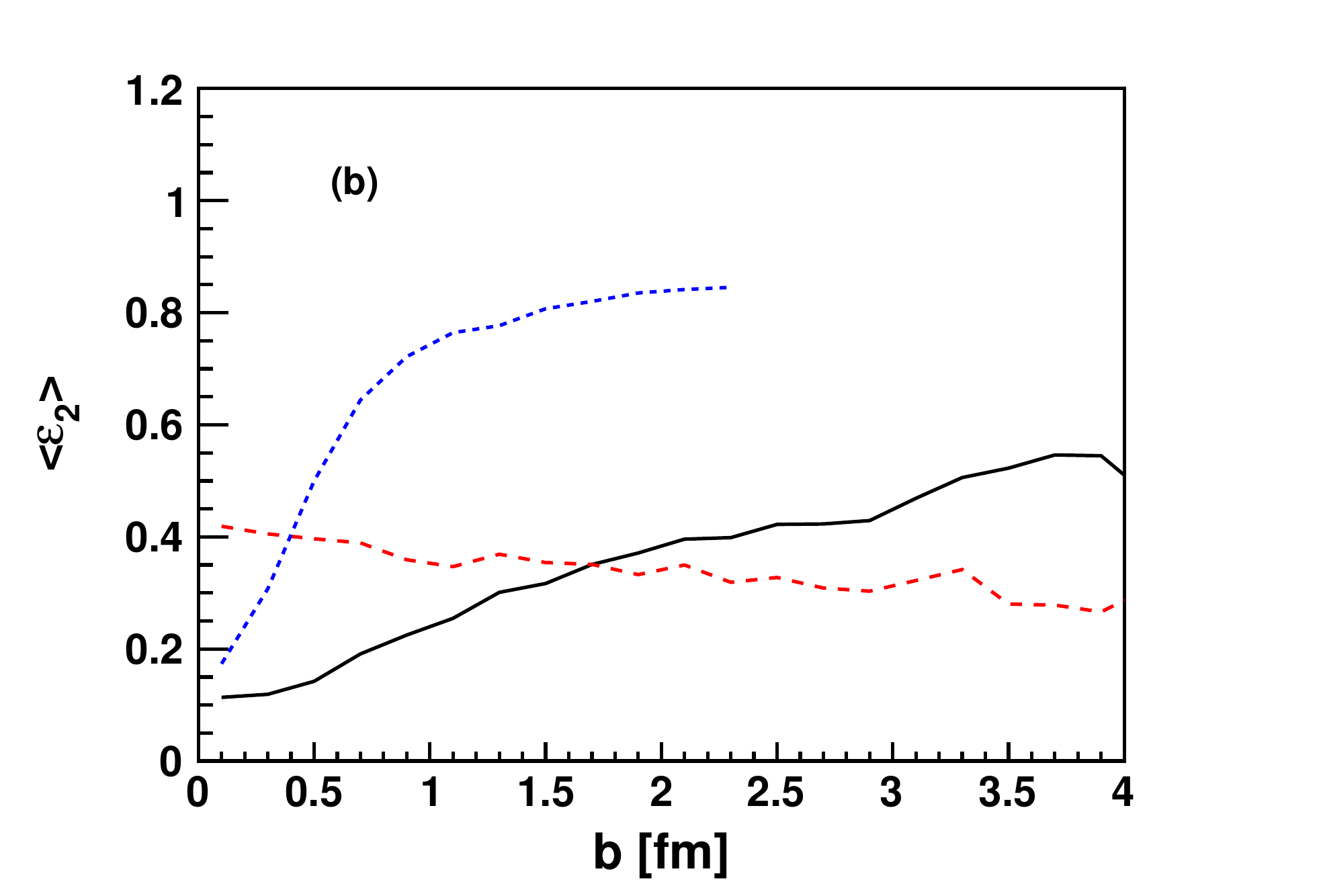}
	\caption{Initial eccentricity versus the number of partons after string melting (a) and impact parameter (b) for proton geometry with overlap function method (black solid line) and quark constituent picture (red dashed line) from our approach in contrast to those from the full AMPT model (blue dotted line) using parton positions right after string melting.}
	\label{fig:ecc2}    
\end{figure*}

We present the initial parton spatial distributions in the transverse plane right after string melting in Fig.~\ref{fig:geo} for different proton geometry assumptions. Only partons with formation time $t_p<5$ fm/c are displayed in this figure. The solid dots represent the initial parton positions at each parton's individual formation time, while the thin arrows indicate their transverse velocity vectors (with the length proportional to the velocity magnitude $\beta_{\perp}=p_{T}/E$). The thick red arrow gives the direction of the second order event plane obtained from the positions of the shown partons. The spatial eccentricities and event plane directions are obtained following the calculations made for participating nucleons in Ref.~\cite{Koop:2015wea}. Partons that will experience rescatterings in ZPC are shown in red and the light green partons have no interactions in parton cascade. The two black circles with radius $R=0.8$ fm demonstrate the transverse view of the proton beams located at $(-b/2, 0)$ and $(b/2, 0)$. The first row represents results obtained with initial conditions based on the $T(x,y,b)$ weighting method and the second row includes the sub-nucleon fluctuations within the constituent quark picture. We find that the transverse spatial structure of the produced initial parton system largely depends on the implemented proton geometry. With $T(x,y,b)$ weights, the partons are mostly distributed around the center of the overlap region for two proton beams at $b=0.1$ fm as shown in Fig.~\ref{fig:geo} (a). In peripheral collisions ($b=0.6$ fm), the parton system becomes stretched along the impact parameter direction in Fig.~\ref{fig:geo} (b). On the other hand, it is observed in Fig.~\ref{fig:geo} (c) and (d) that multiple hotspots or localized high density regions can be produced with the constituent quark picture around the binary collision centers of each interacted quark pair shown by the yellow stars. 

Figure.~\ref{fig:ecc2} shows the average initial eccentricity distributions obtained from the parton spatial positions right after string melting versus the parton number and impact parameter. Similar to Fig.~\ref{fig:geo}, only partons with $t_p<5$ fm/c are considered for the estimation of eccentricity while $N_{parton}$ in Fig.~\ref{fig:ecc2}(a) represents the number of partons in the full phase space melted from the initial primary hadrons in that event. Eccentricities are calculated with the initial positions at each parton's formation time~\cite{Nagle:2017sjv}. For the overlap function weighting method, the eccentricity is largely driven by the geometric shape of the transverse overlap area. The eccentricity drops significantly from peripheral collisions to central collisions. However, the quark constituent picture that has sub-nucleon fluctuations generates larger eccentricities in central collision events in contrast to the overlap function weighting method. The impact parameter dependence of the eccentricities from the quark constituent model is quite weak. We also present the eccentricity distribution from the full AMPT model in Fig.~\ref{fig:ecc2} with the blue dotted line. Its value is found to be significantly larger than that from our current approach except for the events with very small $b$. The origin of the eccentricity in AMPT and how it affects the final state collectivity will be discussed at the end of Sec.~\ref{sec:results}. 

\begin{figure}
	\centering
	\includegraphics[width=0.5\textwidth]{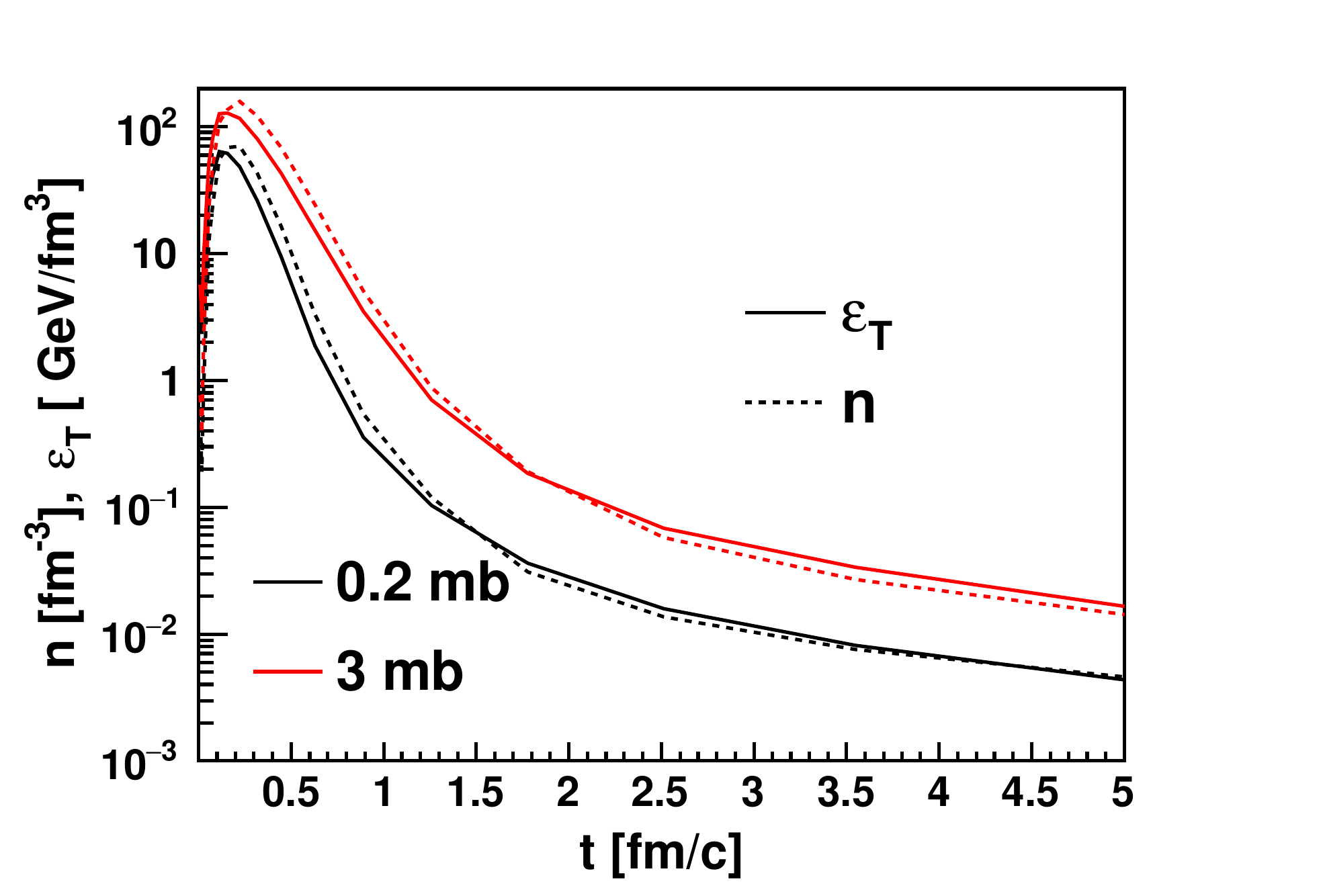}
	\caption{Time evolutions of energy (solid) and number (dashed) density for the parton system generated in pp collisions at $b=0$ fm with parton rescattering cross section 0.2 mb (black) and 3 mb (red) from the overlap function weighting method.}
	\label{fig:evolve}  
\end{figure}

With the overlap function weighting method, time evolutions of the energy density and the number density of the partons are shown in Fig.~\ref{fig:evolve}. In this comparison, the parton densities are obtained with all the active partons which are formed at $t_{p}<t$ and not yet hadronized at time $t$. The density distributions of the parton system in the central cell of pp collisions at $\sqrt{s}=13$ TeV with $b=0$ fm using different final state parton rescattering cross sections are presented. The density of partons decreases at large $t$ due to the expansion of the system. The solid line represents the transverse energy density $\varepsilon_{T}$ and the dashed line shows the parton number density $n$. The central cell is defined with a cross sectional area with the transverse radius of 0.5 fm and a longitudinal dimension covering $-0.5t$ to $0.5t$. The energy density of the system is found to be above the critical energy density ($\sim 1$ GeV/fm$^3$) within $t<1$ fm/$c$, indicating the possibility of producing a short lived QGP droplet in high energy pp systems. By increasing the parton rescattering cross section in ZPC from 0.2 mb to 3 mb, the energy density of the system becomes larger at any given time during the entire evolution stage. This observation implies that the deconfined matter in pp collisions with stronger final state interactions can have a longer lifetime. It is also worthwhile to note that the typical formation time of the partons is around 0.1-0.5 fm/$c$ in the string melting scheme, different from the string fragmentation time at about 1 fm/$c$. As is emphasized in Ref.~\cite{Lin:2004en,Nagle:2017sjv}, the general time scale in the string fragmentation picture may not apply to the initial partons as string melting mechanism models parton not hadron productions from the string energy field.

\section{Results}
\label{sec:results}
The final state particle productions are compared to the pp 13 TeV experimental data after all the secondary interactions and resonance decays are finished in the transport model. It is found in our study that different proton geometries for the initial condition have a negligible impact on the inclusive particle productions. Initial state conditions mainly reveal themselves in the correlation measurements. Therefore, we will show the effects of final state interactions on particle yields and spectra using the constituent quark initial condition in Sec.~\ref{subsec:spectra} and compare different initial conditions on long range two particle correlations in Sec.~\ref{subsec:correlation}. The results from the public AMPT code (v2.26t9)~\cite{ampt_web} are also presented in the comparisons with Lund fragmentation parameters taken as $a=0.5$, $b=0.9$ GeV$^{-2}$ and the parton rescattering cross section set to 1.5 mb.  

\subsection{Particle spectra and final state interactions}
\label{subsec:spectra}
\begin{figure}
	\centering
	\includegraphics[width=0.5\textwidth]{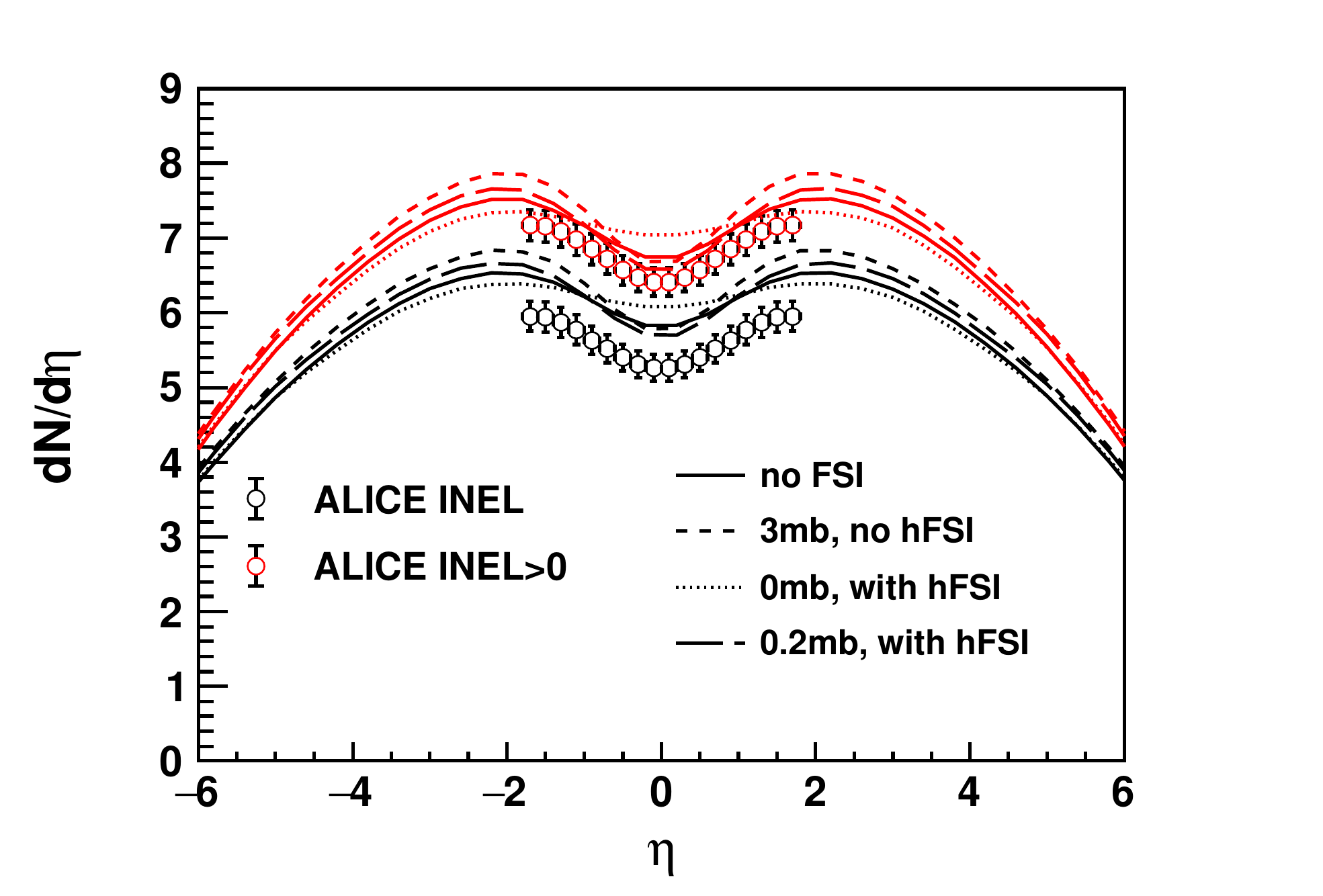}
	\caption{Charged hadron pseudo-rapidity distribution in pp 13 TeV with no final state rescattering (solid line), only parton rescattering with 3mb cross section (dashed line), only hadron rescattering (dotted line) and combined parton/hadron rescatterings (long dashed line) compared to ALICE data in inelastic (black circle) and INEL$>0$ events (red circle)~\cite{Adam:2015pza}. Black lines and red lines are the model calculations obtained with the ALICE INEL and INEL$>0$ cut, respectively.}
	\label{fig:charge_dndeta_FS}       
\end{figure}
\begin{figure*}[hbt!]
	\centering
	\includegraphics[width=0.49\textwidth]{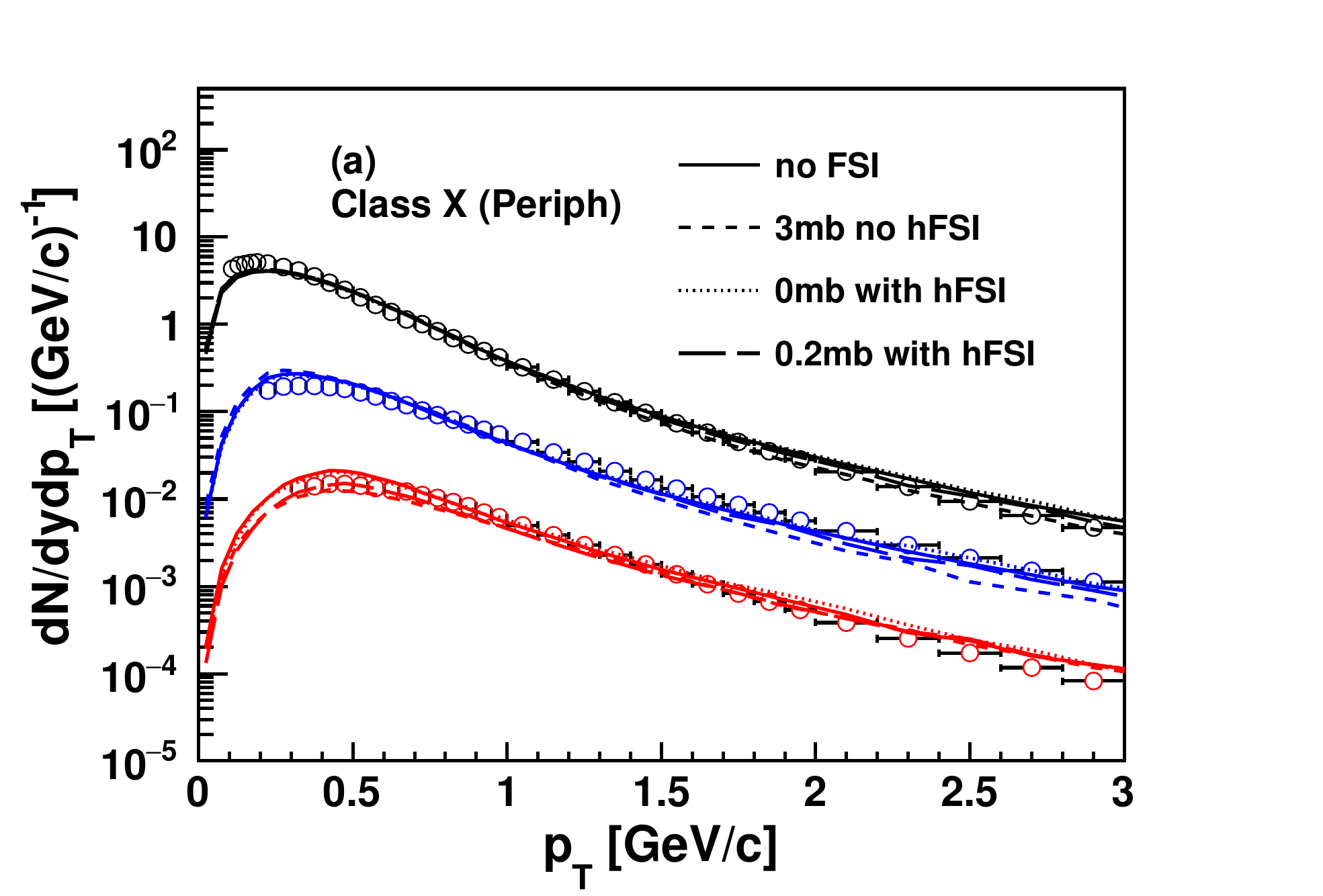}
	\includegraphics[width=0.49\textwidth]{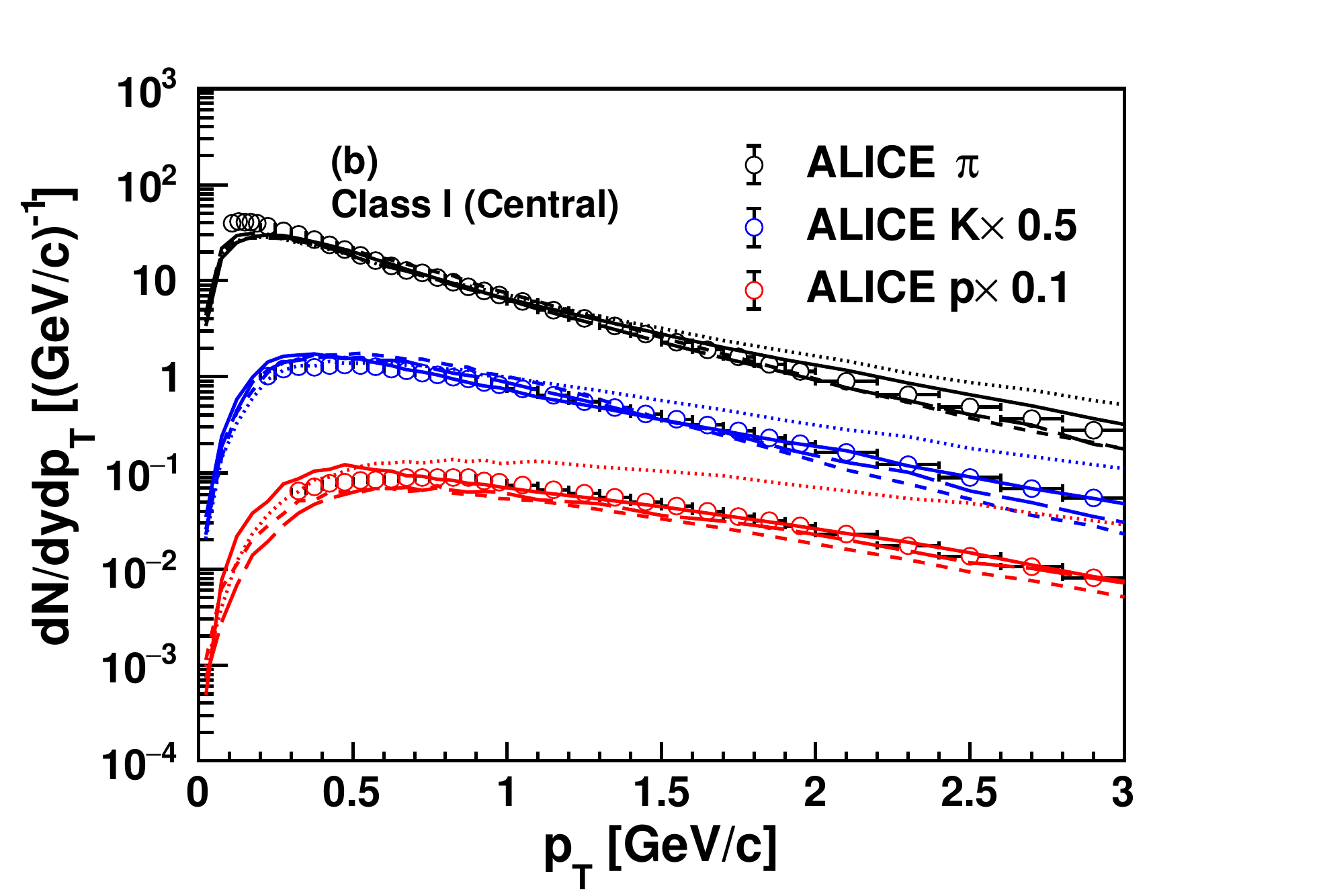}
	\caption{Transverse momentum spectra within $|y|<0.5$ for $\pi$ (black), $K$ (blue) and proton (red) in event class X for peripheral collisions (left) and event class I for central collisions (right) with no final state interactions (solid line), only parton rescatterings (short dashed line), only hadron rescatterings (dotted line) and combined parton/hadron rescatterings (long dashed line). ALICE data are taken from~\cite{Acharya:2020zji}.}
	\label{fig:pikp_cent}    
\end{figure*}
\begin{figure*}[hbt!]
	\centering
	\includegraphics[width=0.49\textwidth]{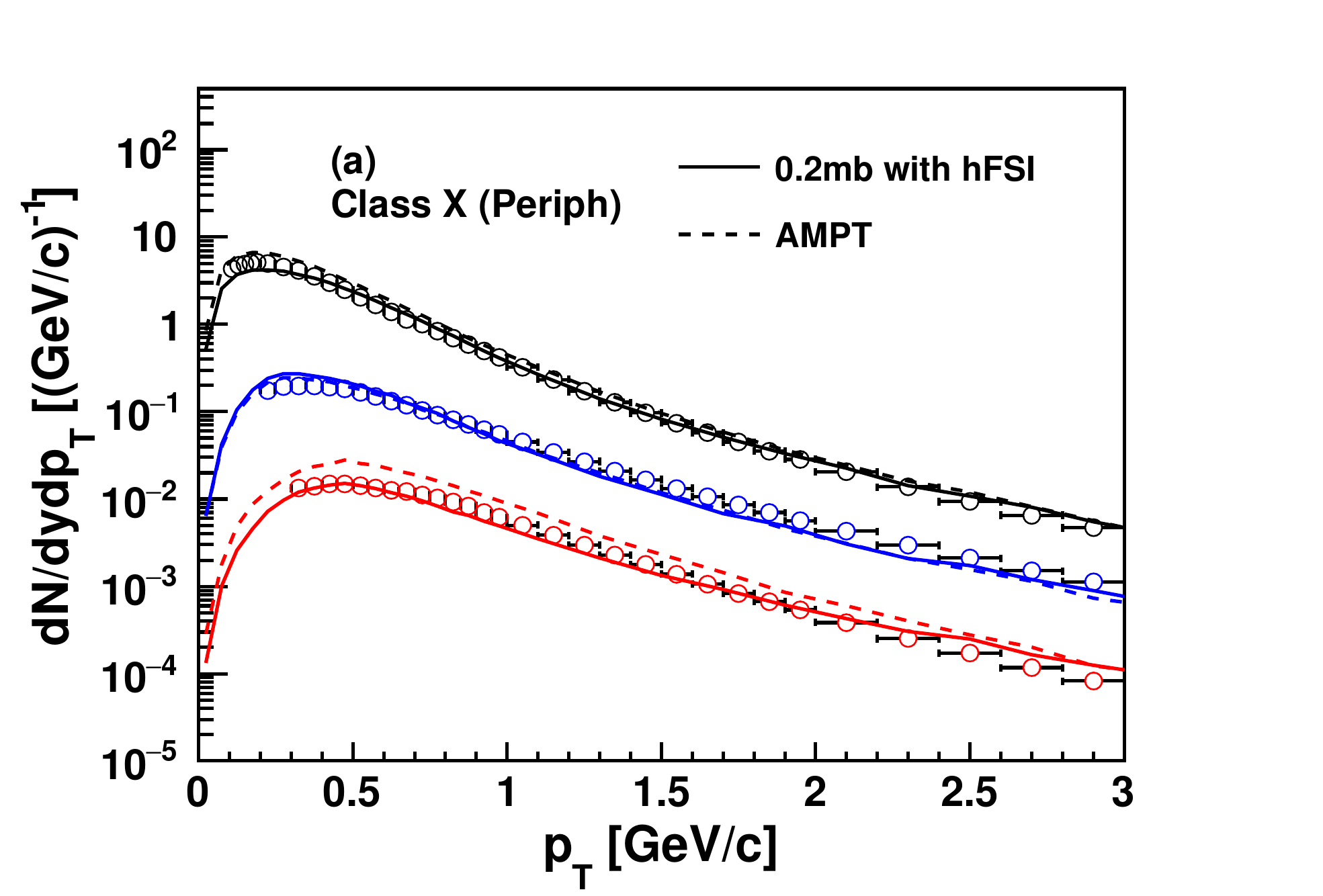}
	\includegraphics[width=0.49\textwidth]{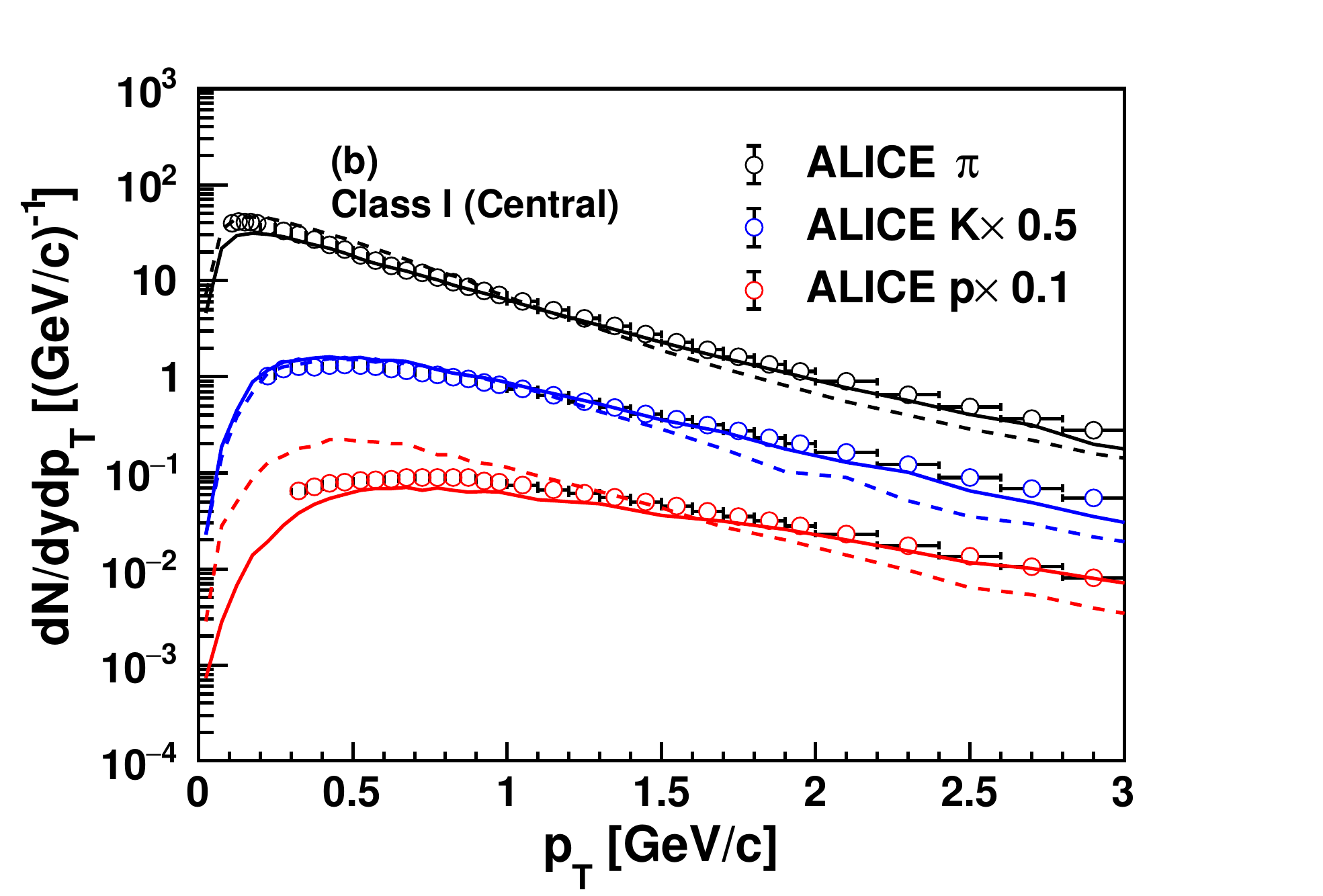}
	\caption{Transverse momentum spectra with $|y|<0.5$ for $\pi$ (black), $K$ (blue) and proton (red) in event class X for peripheral collisions (left) and event class I for central collisions (right) compared to AMPT calculations. Solid lines represent our current calculations with 0.2 mb parton rescattering cross section and final state hadron interactions. Dashed lines show AMPT calculations. ALICE data are taken from~\cite{Acharya:2020zji}.}
	\label{fig:pikp_cent_ampt}    
\end{figure*}

The charged hadron pseudo-rapidity distributions are investigated in Fig.~\ref{fig:charge_dndeta_FS} with different choices on the final state interactions using the constituent quark geometry for the proton. Final state interactions are denoted as FSI and hadronic final state interactions are denoted as hFSI in the figures. Solid lines represent the calculations with both parton and hadron rescatterings switched off. The results with only final state parton interactions or hadron interactions are shown in dashed and dotted lines, respectively. A quite large parton rescattering cross section 3 mb has been used in this comparison to explore the maximized parton evolution effects. Calculations are compared to the minbias measurements from ALICE inelastic events (black markers) and INEL$>$0 events with at least one charged track in mid-rapidity (red markers)~\cite{Adam:2015pza}. It is interesting to observe that, unlike AA collisions, final state interactions do not change the charged multiplicity density too dramatically in pp. As parton rescatterings generally decrease the parton transverse momentum, it is observed that the mid-rapidity particles are shifted to forward and backward $\eta$ regions by comparing the solid and dashed lines. On the other hand, the transition from solid to dotted lines is an indication of the radial flow effect induced by the hadronic rescatterings, which causes the excess in the charged particle density at $\eta\sim 0$. Stronger radial expansion arised from hadron rescatterings increases the average $p_T$ of charged hadrons and pushes more particles to the mid-rapidity region. It is also found that the long dahsed line, with combined parton and hadron final state effects and a moderate parton rescattering cross section 0.2 mb, reproduces the INEL$>$0 charged particle density data reasonably, although the inelastic event charged particle yields are overestimated.

In Fig.~\ref{fig:pikp_cent}, we compare the transverse momentum spectra for $\pi^{\pm}$, $K^{\pm}$ and $p(\bar{p})$ in our model to the ALICE pp data at $ \sqrt{s}=13$ TeV~\cite{Acharya:2020zji} in different centralities. The percentile definition of the events are made with the charged track number in the pseudo-rapidity regions $-3.7<\eta<-1.7$ and $2.8<\eta<5.1$ following the event classification scheme used at the ALICE experiment. Ten event classes (Class I $\sim$ Class X from central to peripheral) are obtained with the same analysis procedure in~\cite{Acharya:2020zji}. The results for the most peripheral event class 64.5-100\% (Class X) and the most central event class 0-0.92\% (Class I) can be found in Fig.~\ref{fig:pikp_cent} (a) and (b), respectively. Similar to what has been done in Fig.~\ref{fig:charge_dndeta_FS}, we only show the comparisons of final state effects based on the constituent quark proton matter distribution. Considering that the string melting framework generally works better for low to intermediate $p_T$ region, we focus on the $p_T$ range from 0 to 3 GeV$/c$ in the spectra comparisons. Results without final state interactions are presented in solid lines. Short dashed and dotted lines represent the calculations with only parton rescatterings (3 mb in ZPC, ART turned off) and only final hadron interactions (0 mb in ZPC, ART turned on), respectively. The long dashed lines represent the calculations that include both parton rescatterings and final state hadron interactions. 

\begin{figure*}
	\centering
	\includegraphics[width=0.49\textwidth]{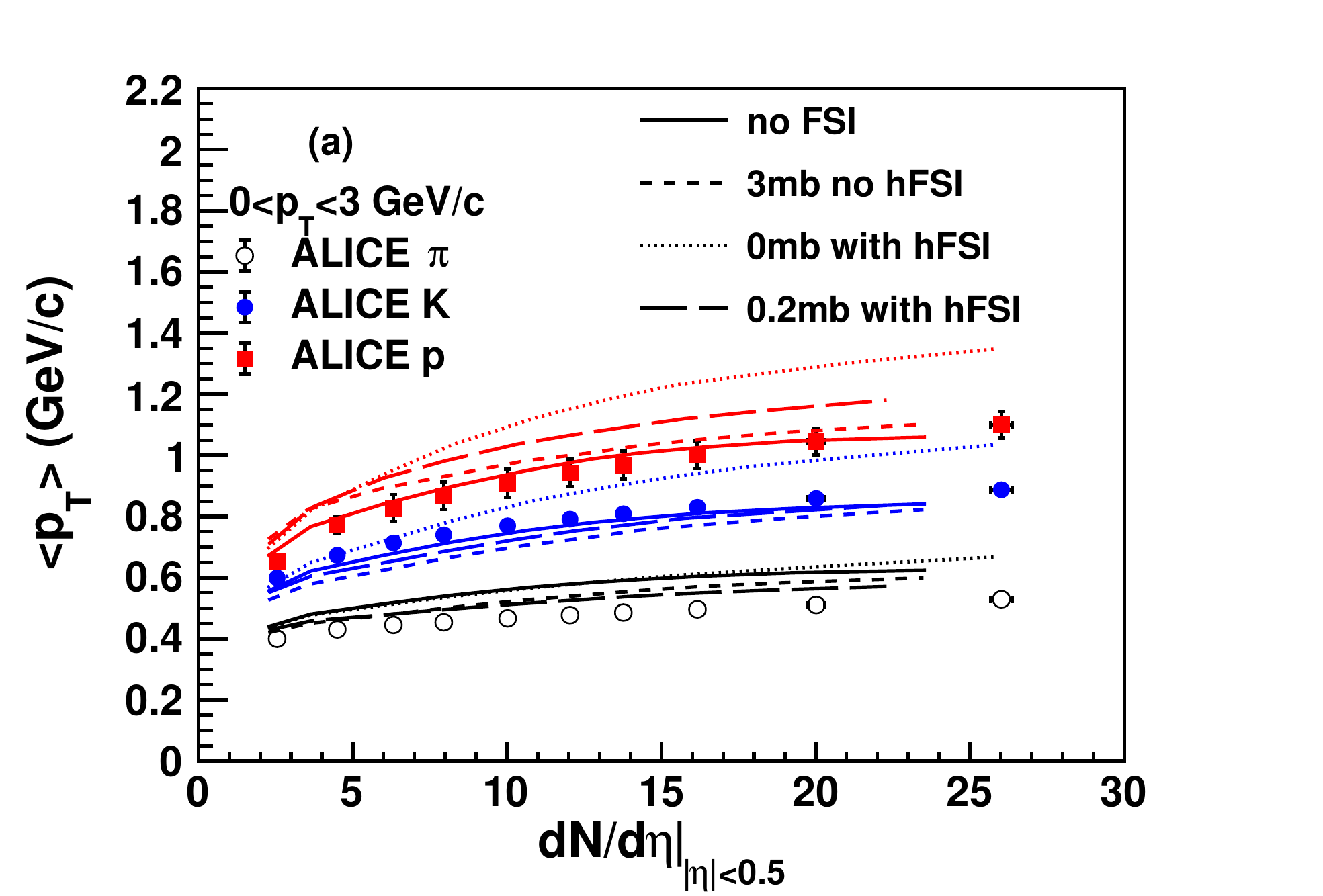}
	\includegraphics[width=0.49\textwidth]{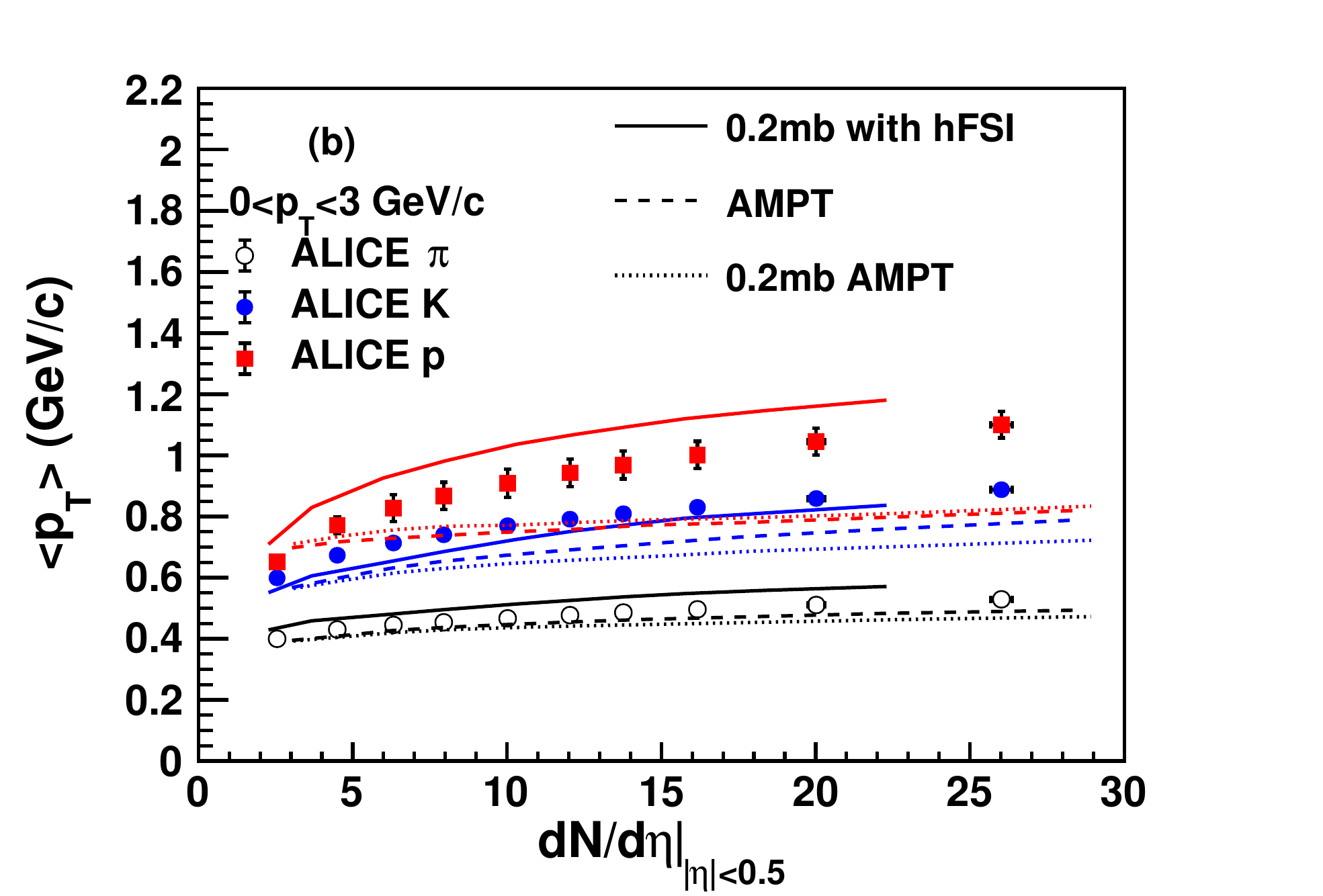}
	\caption{Average transverse momentum of $\pi$ (black), $K$ (blue) and proton (red)  at mid-rapidity with $0<p_T<3$ GeV$/c$ versus the charged hadron multiplicity density. Final state effects are shown in (a). Comparisons to AMPT are shown in (b). Central values of ALICE data are obtained with a refit to the data from~\cite{Acharya:2020zji}.}
	\label{fig:Avgpt_pikp}
\end{figure*}

\begin{figure*}[h!]
	\centering
	\includegraphics[width=0.49\textwidth]{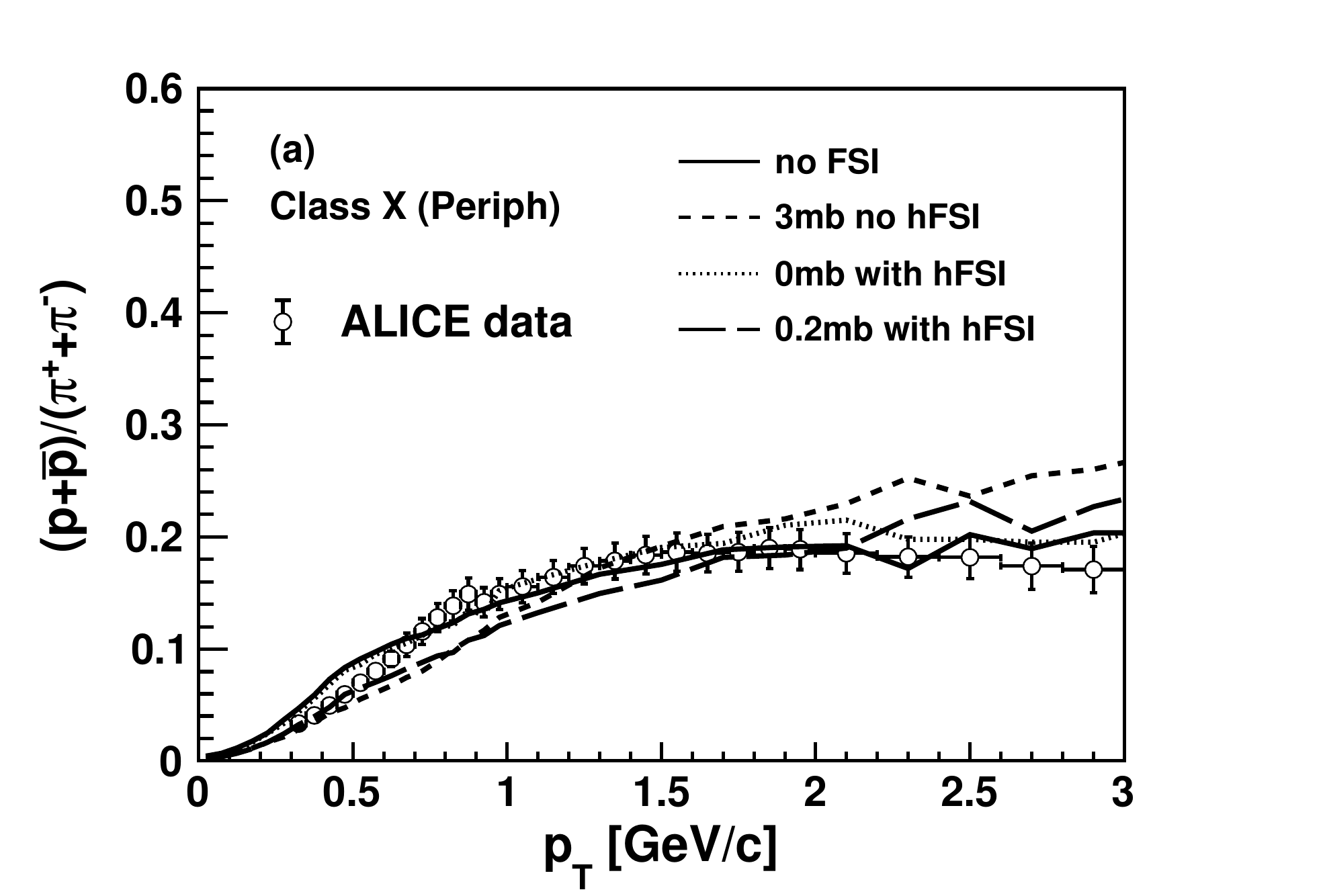}
	\includegraphics[width=0.49\textwidth]{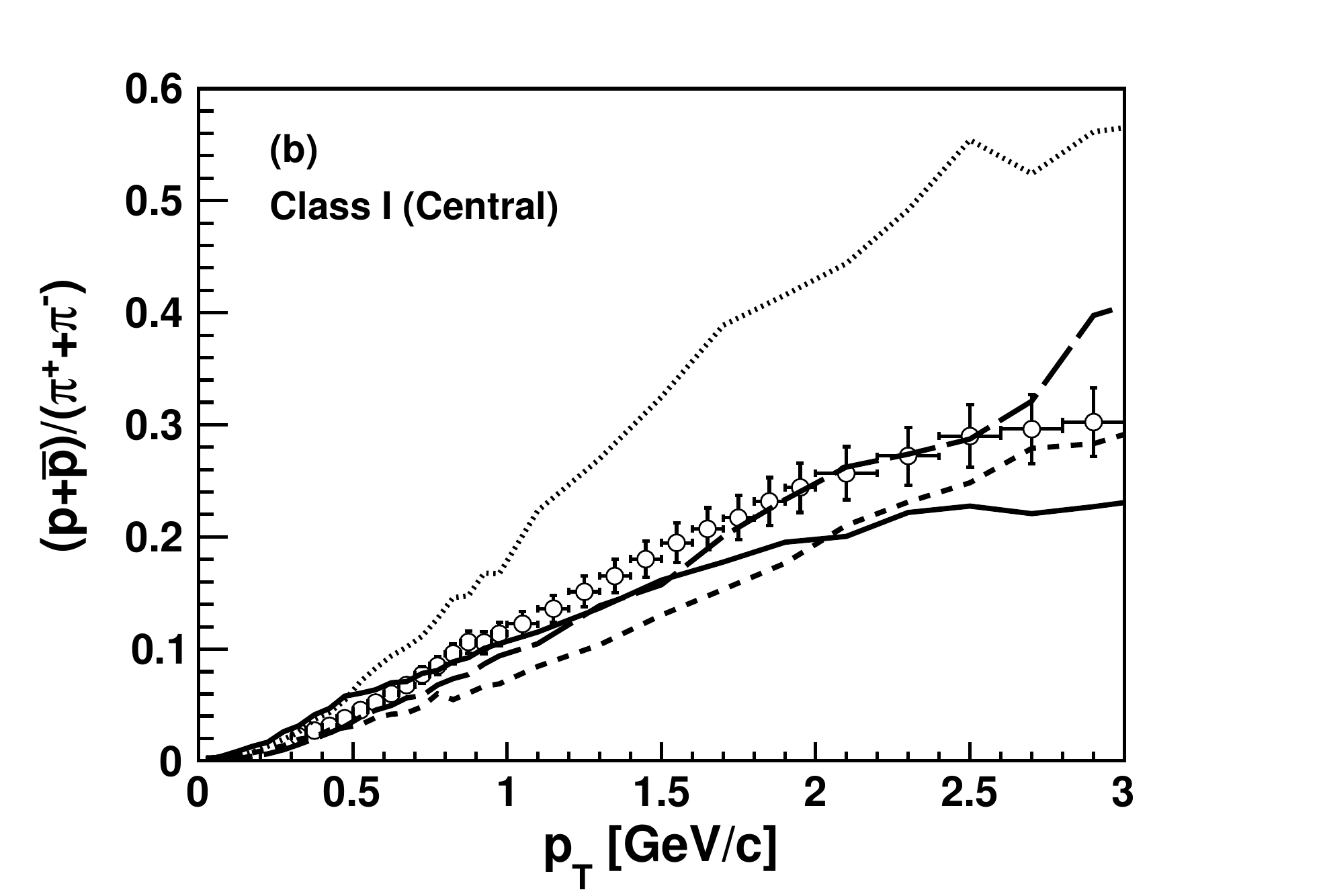}
	\caption{$p_T$-differential $p/\pi$ ratio at mid-rapidity in event class X for peripheral collisions (a) and in event class I for central collisions (b) with no final state interactions (solid line), only parton rescatterings (short dashed line), only hadron rescatterings (dotted line) and combined parton/hadron rescatterings (long dashed line). Data are taken from~\cite{Acharya:2020zji}.}
	\label{fig:POverPi_pt_FSI}
\end{figure*}
In the peripheral event class with low multiplicities shown in Fig.~\ref{fig:pikp_cent} (a), the final state interaction effects are expected to be very weak. Differences between various final state interaction combinations are indeed found to be small. In central pp collisions shown in Fig.~\ref{fig:pikp_cent} (b), a significant stiffening can be observed for the $p_T$ spectra with only hadron rescatterings effects (dotted line). The scenario of pure hadronic rescatterings has a strong radial expansion, boosting the hadrons to higher $p_T$ with a mass ordering feature. Protons receive stronger change of $p_T$ compared to the lighter $K^{\pm}$ and $\pi^{\pm}$. On the other hand, parton rescatterings are expected to decrease the hadron yield at larger $p_T$. When parton and hadron interactions are considered together, the strong radial expansion observed in the pure hadronic rescattering scenario is no longer there (long dashed line) because the parton stage delays hadronic rescatterings to lower densities. Underestimations in high $p_T$ region exist for $\pi^{\pm}$ and $K^{\pm}$ but not for $p(\bar{p})$ since the proton transverse momentum suppression due to parton rescatterings is compensated by the stronger radial flow from secondary hadronic interactions.

We also compare our current results on the multiplicity dependent transverse momentum spectra to the calculations from the public AMPT code in Fig.~\ref{fig:pikp_cent_ampt}. In peripheral pp collisions, only the proton yield is slightly overestimated in AMPT and the meson $p_T$ spectra are generally well reproduced. In central pp events, the AMPT model predicts softer transverse momentum spectra especially for protons. Compared to the full AMPT model, our current approach improves the description of the identified particle $p_T$ spectra in high multiplicity events.

Figure~\ref{fig:Avgpt_pikp} (a) shows the final state effects on the multiplicity dependent average transverse momentum distributions for $\pi$, $K$ and $p(\bar{p})$. Only hadrons with $0<p_{T}<3$ GeV$/c$ at mid-rapidity are considered in this comparison. The ALICE $\langle p_T\rangle$ data are obtained by refitting the measured $p_T$ spectra with the L\'{e}vy-Tsallis parameterization restricted to the corresponding $p_{T}$ range while the uncertainties are taken from the experimental data with full $p_T$ range\cite{Acharya:2020zji} directly. Similar to the observation in Fig.~\ref{fig:pikp_cent}, the scenario of pure hadronic rescatterings shown by dotted lines leads to a strong enhancement in the average $p_T$. The magnitude of the enhancement is mass dependent and more pronounced for protons. The scenario of pure parton rescatterings (short dahsed lines) suppresses $\langle p_{T} \rangle$ for pion and kaon but slightly enlarges proton $\langle p_{T} \rangle$. 
Figure.~\ref{fig:Avgpt_pikp} (b) shows a comparison of our current model calculations with all final state interactions to the full AMPT results together with the experimental data of $\langle p_{T} \rangle$. By including both the parton and hadron evolutions, results from our current model calculations reproduce the trend of the data within the selected $p_T$ range, while overestimations are found for $\pi$, $p(\bar{p})$ average $p_T$ and kaon $\langle p_T\rangle$ is slightly smaller than data. The full AMPT calculations reasonably describe the pion $\langle p_T\rangle$ but generally deliver very weak multiplicity dependence for kaon and proton. By decreasing the parton rescattering cross section in the full AMPT from 1.5 mb to 0.2 mb, it is found that the average $p_{T}$ only slightly changes and the flat multiplicity dependence persists. The improvement compared to the full AMPT model on the multiplicity dependence of kaon and proton $\langle p_T\rangle$ presumably comes from the MPI mechanism~\cite{Sjostrand:1987su,Sjostrand:2004pf} implemented with the PYTHIA initial conditions incorporated to our transport model framework.

We present a group of comparisons of the final state interactions on the baryon to meson ratio $p/\pi$ in Fig.~\ref{fig:POverPi_pt_FSI}. Differences among the various final state effects are quite small in peripheral collisions as shown in Fig.~\ref{fig:POverPi_pt_FSI} (a). We see slight depletion in the low $p_T$ and enhancement in the high $p_T$ regime of the $p/\pi$ ratio when parton evolutions are included. The impacts of final state rescatterings are more obvious in central collisions shown in Fig.~\ref{fig:POverPi_pt_FSI} (b). Without any final rescattering effects, the $p/\pi$ ratio at low $p_T$ qualitatively agrees with the data but undershoots in the higher $p_T$ region, missing the enhancement feature from peripheral to central collisions suggested by data. The scenario of pure hadron rescatterings as represented by the dotted line shows a large enhancement of the $p/\pi$ ratio in central pp collisions. The parton rescattering effect seems to be quite similar to that in peripheral collisions. An overall agreement can be reached after both the parton and hadron interactions are included. It is seen in this comparison that both the final state parton and hadron effects are necessary to reproduce the multiplicity dependent baryon to meson enhancement in our transport model approach.

\subsection{Long range correlation and initial state conditions}
\label{subsec:correlation}
\begin{figure*}[hbt!]
	\centering
		\includegraphics[width=0.45\textwidth]{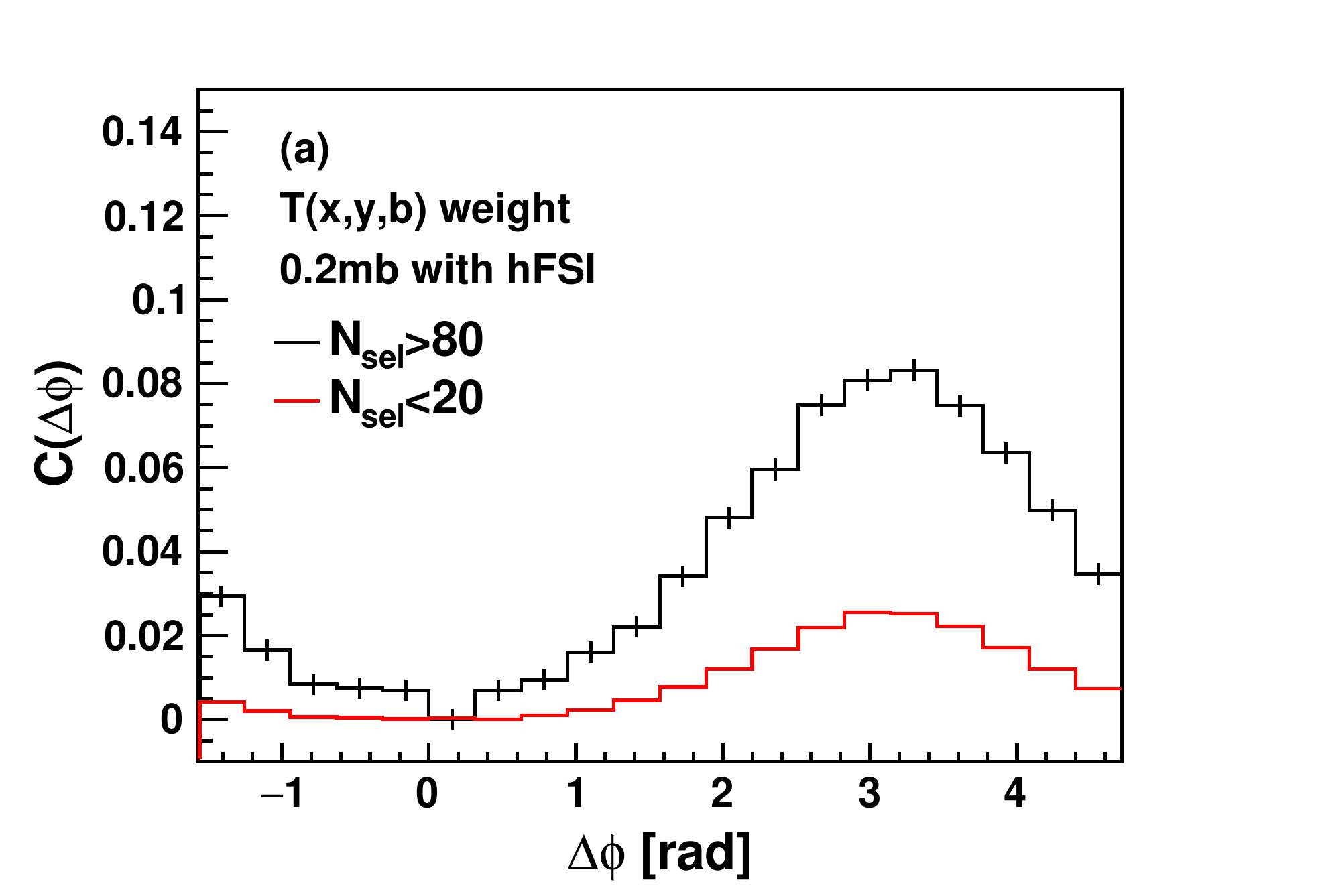}
		\includegraphics[width=0.45\textwidth]{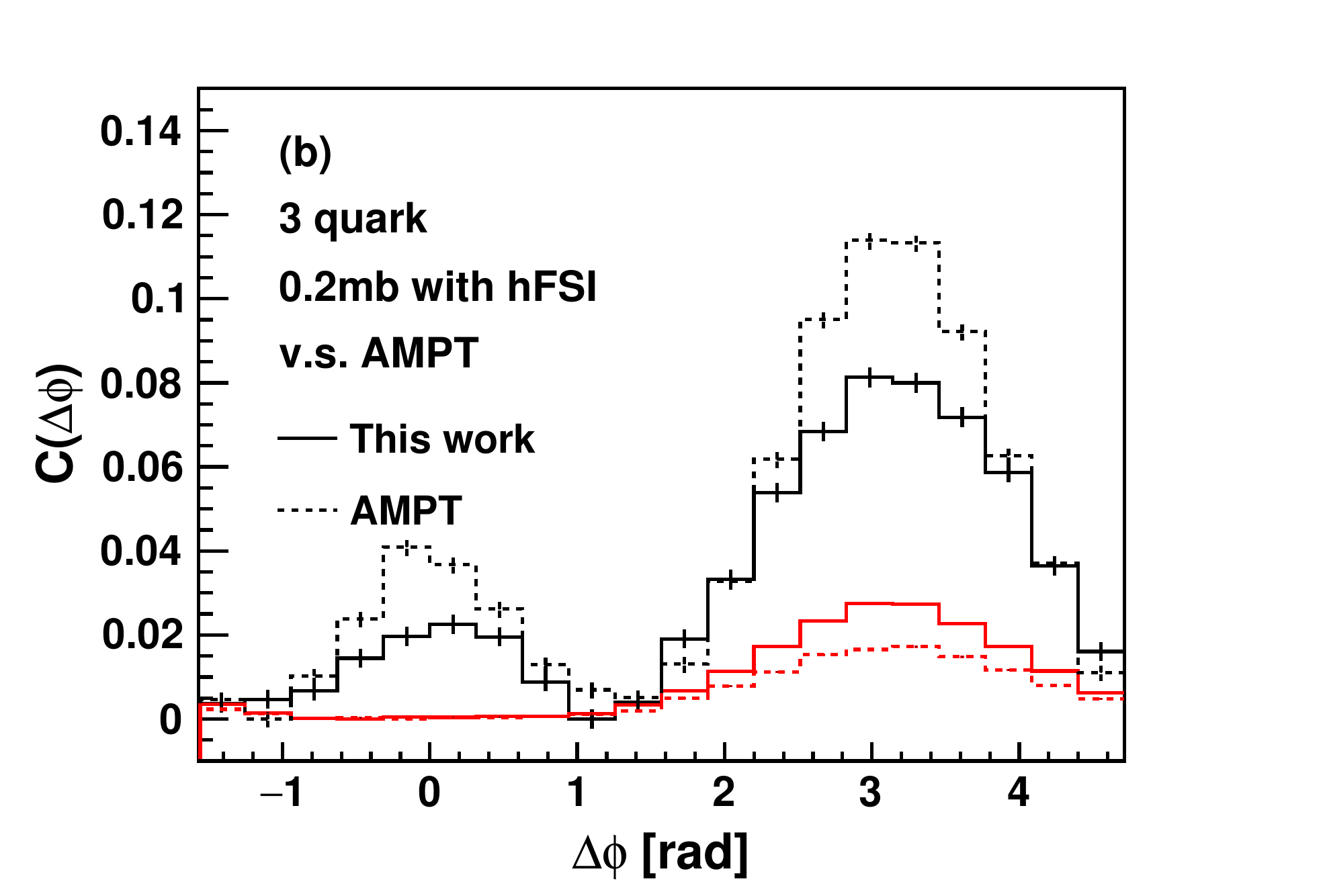}
		\includegraphics[width=0.45\textwidth]{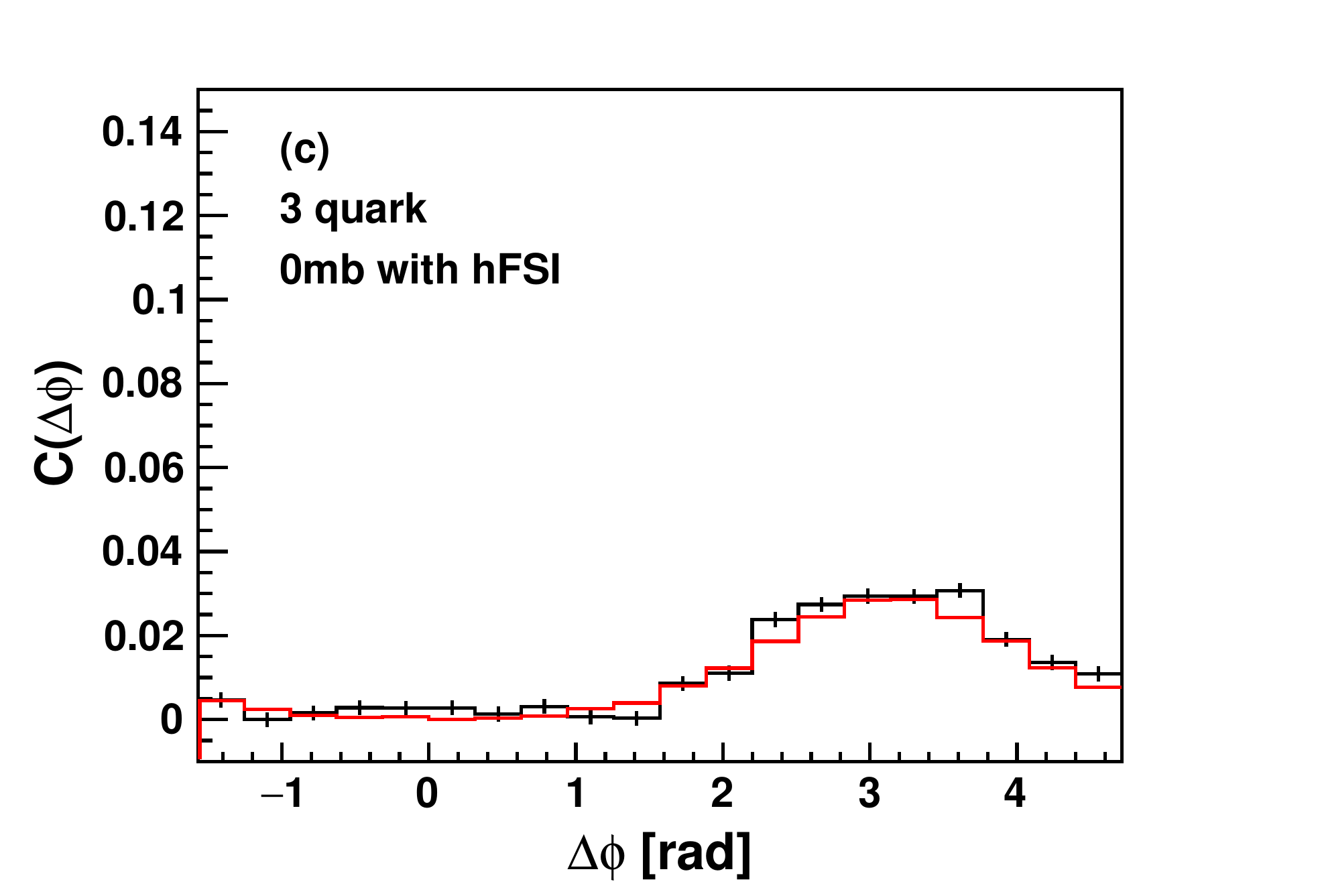}
		\includegraphics[width=0.45\textwidth]{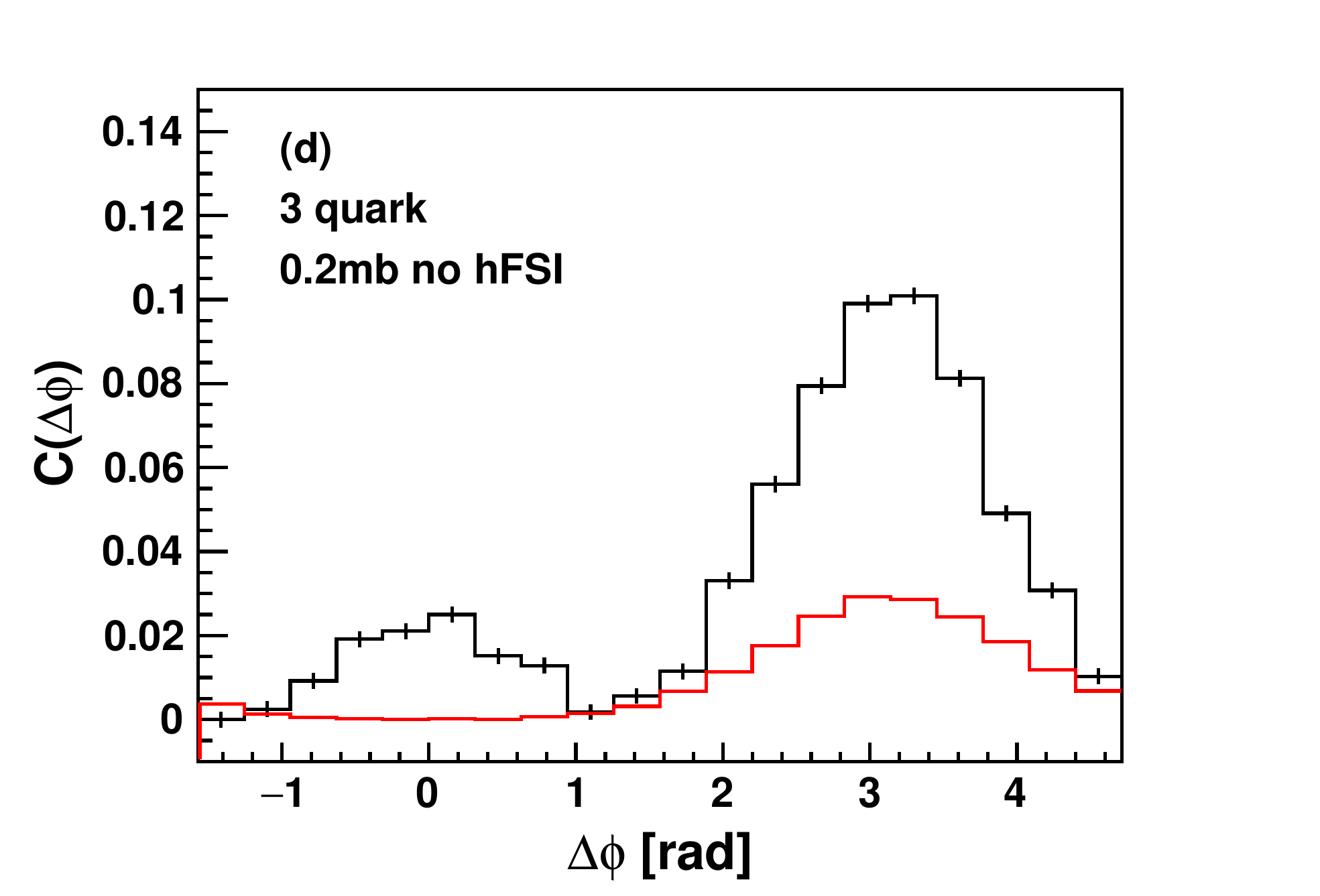}
	\caption{Two particle correlations with large pseudo-rapidity gap $|\Delta\eta|>2$ and $1<p_{T}<3$ GeV$/c$ in high multiplicity events (black line) and low multiplicity events (red line) for initial proton geometry with overlap function method (a) and quark constituent picture (b). The final state effects on the correlation function with only hadron rescatterings (c) and parton rescatterings (d) are also shown based on the quark constituent picture. The corresponding results from full AMPT (dashed line) are shown in (b) for comparison.}
	\label{fig:corre_HM}    
\end{figure*}
Other than the radial flow feature discussed above, particle correlation measurement is a more differential observable to investigate the initial state geometry related collectivity effects in pp collisions. We find that the appearance of the near side ridge structure in two particle long range correlations can be largely related to the initial sub-nucleon fluctuations in transverse space. 

In Fig.~\ref{fig:corre_HM}, we present the multiplicity dependence of the projected correlation functions $C(\Delta\phi)=\frac{1}{N_{trig}}\frac{dN^{pair}}{d\Delta\phi}$ obtained with the overlap function $T(x,y,b)$ weighting method (Fig.~\ref{fig:corre_HM}(a)) and the constituent quark assumption (Fig.~\ref{fig:corre_HM}(b)). The final state parton scattering cross section is 0.2 mb and hadronic rescatterings are switched on in this comparison. The trigger and associate hadrons are selected with the transverse momentum requirement $1<p_{T}<3$ GeV/$c$ in the acceptance of $|\eta|<2.4$ following the analysis performed at the CMS experiment~\cite{Khachatryan:2016txc}. Events are separated into two categories based on $N_{sel}$, the number of selected charge tracks with minimum transverse momentum $p_T>0.4$ GeV/$c$ within $|\eta|<2.4$. High multiplicity events are required to have $N_{sel}>80$ and low multiplicity events are defined with $N_{sel}<20$. The two hadrons in each pair must be separated with a pseudo-rapidity gap $|\Delta\eta|>2$. The correlation function has been corrected following the standard zero-yield-at-minimum procedure. It is suggested by the experimental data that a significant near side ridge structure exists in the correlation function as a local maximum at $\Delta\phi\sim 0$ in the high multiplicity pp events~\cite{Khachatryan:2010gv,Khachatryan:2016txc}. Our study shows a long-range ridge-like structure is present only if the proton matter distribution is modeled based on the constituent quark picture, indicated by the near side peak of black dots in Fig.~\ref{fig:corre_HM}(b). Considering that all the final state interaction parameters are set to be the same in this comparison, this is a quite striking result implying the connection of the induced long range correlation with the underlying sub-nucleon fluctuation effects. 
\begin{figure*}[hbt!]
	\centering
	\includegraphics[width=0.45\textwidth]{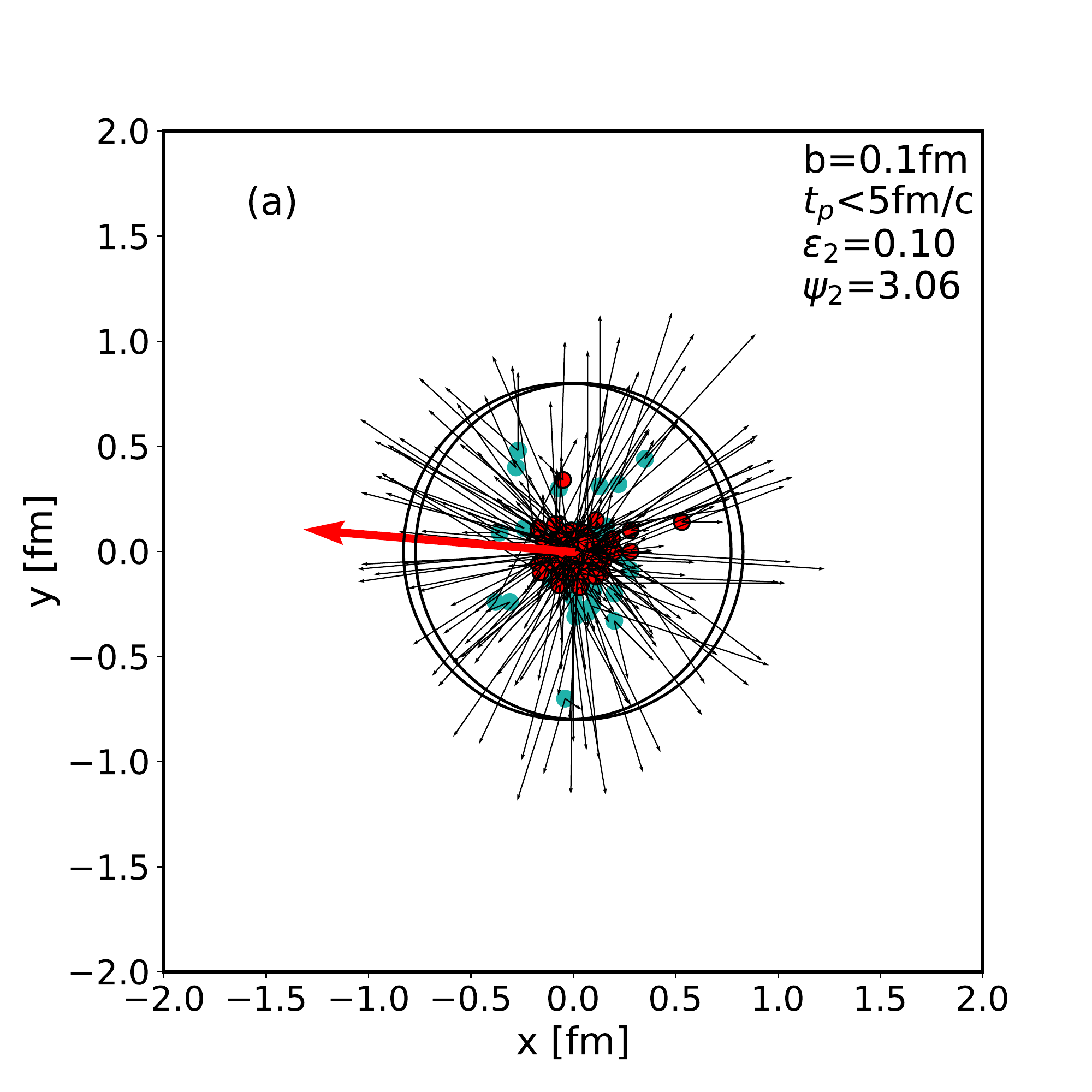}
	\includegraphics[width=0.45\textwidth]{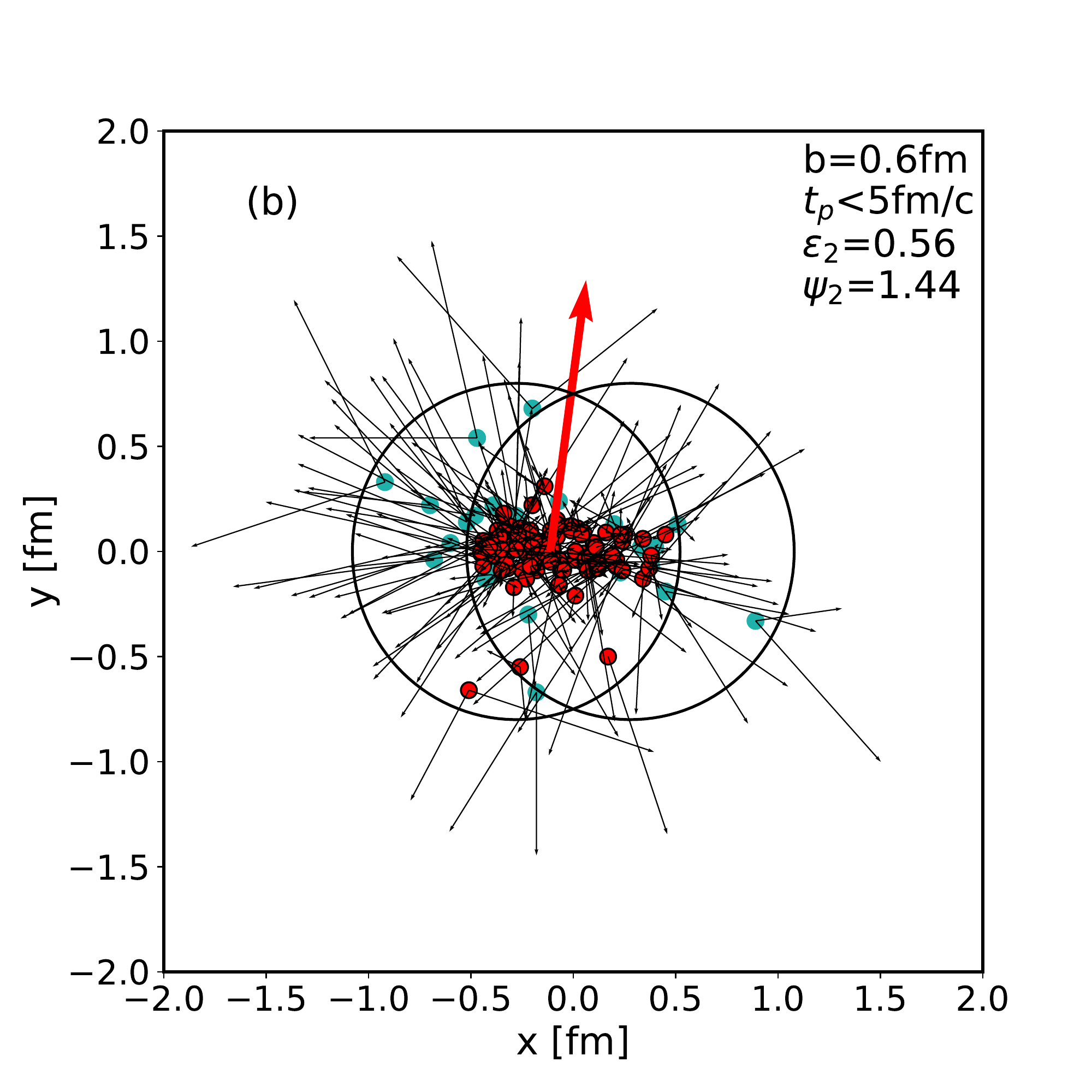}
	\caption{Transverse view of the created parton system from the full AMPT in a central collision at $b=0.1$ fm (a) and in a peripheral collision at $b=0.6$ fm (b). Coordinates of the partons are represented by the dots. Initial parton velocity vectors are shown as black thin arrows.  }
	\label{fig:ampt_geo}       
\end{figure*}

%\begin{figure}[hbt!]
%	\centering
%	\includegraphics[width=0.45\textwidth]{dphi_ampt.pdf}
%	\caption{Two particle correlations with large pseudo-rapidity gap $|\Delta\eta|>2$ and $1<p_{T}<3$ GeV$/c$ in high multiplicity events (black line) and low multiplicity events (red line) from the full AMPT model. }
%	\label{fig:ampt_corre}    
%\end{figure}

%\begin{figure}[hbt!]
%	\centering
%	\includegraphics[width=5.91cm]{fig_trans_view_ampt.jpg}
%	\caption{An illustration of the spatial distribution of the excited strings in the transverse plane from the full AMPT model for pp collisions. }
%	\label{fig:trans_space_ampt}  
%\end{figure}

We also ran the same analysis by considering only hadronic rescatterings (0mb parton rescattering cross section) and only parton rescatterings (0.2mb cross section without hFSI) based on the quark constituent assumption in Fig.~\ref{fig:corre_HM}(c) and Fig.~\ref{fig:corre_HM}(d), respectively. The near side ridge persists at high multiplicity when the parton rescattering is included. No near side peak can be found in high multiplicity events with only final hadronic interactions. Therefore, the near side ridge like structure in correlation function is more likely to be developed in the parton evolution stage.

It is interesting to see that the long range near side correlations also exist within the full AMPT model but from a different perspective. Large spatial eccentricity can be induced via the three fire ball like arrangement along the impact parameter direction in AMPT as long as the impact parameter $b$ is not too small to eliminate the separation of different particle sources as already shown in Fig.~\ref{fig:ecc2}. The spatial profile of the initial partons after string melting from the full AMPT model is presented in Fig.~\ref{fig:ampt_geo}. We show the velocity and position of the partons with $t_p<5$ fm$/c$ at their formation time with impact parameter b=0.1 fm and b=0.6 fm. For central collisions, the partons are roughly placed around the same position. The parton spatial distribution is elongated on the impact parameter direction for peripheral events. Figure~\ref{fig:corre_HM}(b) also shows the long range correlation function from the full AMPT model in the dash lines, wherein the near side ridge arises in high multiplicity pp events due to the parton evolution effects developed from highly eccentric initial geometries~\cite{Ma:2014pva,Nagle:2017sjv}. 

We believe that the sizable initial eccentricities created in our current work and the full AMPT model are the key factors to reproduce the flow like long range correlations found in high multiplicity pp data. The separated particle sources in the transverse plane are necessary to generate the large eccentricities for the evolving parton system within the transport model framework. In our current work, the spatial separation is produced through the hot spots like regions due to the sub-nucleon fluctuations using quark constituent assumptions. The full AMPT does not include any sub-nucleon structures but models the initial parton spatial distribution with two or three fire balls along the impact parameter. As the string shoving mechanism also generates significant long range two particle correlations, more detailed studies to understand the implications of these different model implementations are planned in our future work.

\section{Summary}
\label{sec:summary}
Experimental results revealing collectivity like behaviors in high multiplicity pp collisions have been considered as evidence supporting the creation of deconfined quark gluon medium in small systems. In this work, we introduce a novel transport model approach to systematically study the collective phenomena observed in pp events. We combine PYTHIA8 initial states and AMPT final state interactions together with several options on the proton geometry assumptions in this approach. We show in the study that both parton and hadron final state interactions are important to understand the multiplicity dependent mass ordering in the pp transverse momentum spectra. The near side ridge structure observed in two hadron long range correlations is found to be developed during the partonic phase in our transport model approach. This observation can be regarded as an indication for the creation of deconfined quark matter in high multiplicity pp events. Our study also shows that the appearance of the near side ridge is affected by the proton sub-nucleon fluctuations. 

We also note that final state rescattering for pp collisions focused on the hadronic interaction channels has been implemented within the PYTHIA event generator itself~\cite{Sjostrand:2020gyg}. Different space-time structure has been established from the string fragmentation perspective~\cite{Ferreres-Sole:2018vgo}, whereas sizable hadronic interaction effects are also observed. It is thus worthwhile to extend the application of our current approach to pA and AA within the Angantyr formalism implemented in PYTHIA8 framework. Examples like this including using UrQMD or the PYTHIA internal hadronic interaction model to handle hadron rescatterings based on the PYTHIA Angantyr space time picture have been extensively applied to describe pA and AA collisions~\cite{daSilva:2020cyn,Bierlich:2021poz}. In contrast, our study has the capability of considering parton evolution effects in a coherent picture. The development of this approach will lay a solid foundation for future studies of various mechanisms, such as string shoving and parton/hadron evolutions, within the same model and help us understand the origin of the collective phenomena in pp collisions.

\begin{acknowledgement}
We are grateful to Qing Wang, Benhao Sa, Shusu Shi, Yuliang Yan, Chao Zhang and Daimei Zhou for helpful discussions. This work is supported by the National Natural Science Foundation of China (Nos. 11905188, 11975078, 11875143), the National Science Foundation under Grant No. 2012947 (Z.-W. L.) and the Innovation Fund of Key Laboratory of Quark and Leption Physics LPL2020P01 (LZ). 
\end{acknowledgement}

\bibliographystyle{epjstyle}   
\bibliography{reference}   % name your BibTeX data base

\end{document}